\documentclass[journal]{IEEEtran}
\usepackage{amsfonts}
\usepackage{amssymb}
\usepackage{amsmath}
\usepackage{mathrsfs}
\usepackage{graphicx,psfrag}
\usepackage{amsxtra}
\usepackage{cite}
\usepackage{setspace}
\usepackage{algorithm}
\usepackage{algorithmic}
\usepackage{multirow}
\usepackage{subeqnarray}
\usepackage{subfigure}
\usepackage{color}
\usepackage{epstopdf}

\begin{document}
\bibliographystyle{unsrt}
%
\title{Backhaul Limited Asymmetric Cooperation for MIMO Cellular Networks via Semidefinite Relaxation}
%
%
%

\author{Fuxin Zhuang,~\IEEEmembership{Student Member,~IEEE,}
        Vincent K.N. Lau,~\IEEEmembership{Fellow,~IEEE.}
\thanks{The authors are with the Department
of Electrical and Computer Engineering, Hong Kong University of Science and Technology, Clear Water bay,
Kowloon, Hong Kong E-mail: fzhuangaa@ust.hk, eeknlau@ust.hk.}
}

%
%
\maketitle

\begin{abstract}
Multicell cooperation has recently attracted tremendous attention because of its ability to eliminate intercell interference and increase spectral efficiency. However, the enormous amount of information being exchanged, including channel state information and user data, over backhaul links may deteriorate the network performance in a realistic system. This paper adopts a backhaul cost metric that considers the number of active directional cooperation links, which gives a first order measurement of the backhaul loading required in asymmetric Multiple-Input Multiple-Output (MIMO) cooperation. We focus on a downlink scenario for multi-antenna base stations and single-antenna mobile stations. The design problem is minimizing the number of active directional cooperation links and jointly optimizing the beamforming vectors among the cooperative BSs subject to signal-to-interference-and-noise-ratio (SINR) constraints at the mobile station. This problem is non-convex and solving it requires combinatorial search. A practical algorithm based on smooth approximation and semidefinite relaxation is proposed to solve the combinatorial problem efficiently. We show that semidefinite relaxation is tight with probability 1 in our algorithm and stationary convergence is guaranteed. Simulation results show the saving of backhaul cost and power consumption is notable compared with several baseline schemes and its effectiveness is demonstrated.
\end{abstract}


%
\IEEEpeerreviewmaketitle

\section{Introduction}
Multicell cooperation (MCP) is a promising technique, which can be used to dramatically improve the performance of cellular networks by coordinating the base stations (BSs) through high speed backhaul links \cite{Gesbert2010}. User data and channel state information (CSI) can be shared over the backhaul links to jointly encode and transmit data signals to users with multicell cooperative processing in order to exploit the intercell interference, which is known to be a limiting factor in conventional cellular networks.

Despite the great potential in combating intercell interference, the enormous signaling overhead incurred during the information exchange in the backhaul links is quite unrealistic in a practical system, where the backhaul links have finite capacity \cite{sanderovich2009uplink,simeone2009downlink}. There are several existing approaches in the literature to cope with this limitation. Interference coordination is introduced to combat intercell interference by exchanging CSI (instead of data payload), and hence, it can significantly reduce the backhaul loading. For example, multicell joint scheduling, power control and beamforming have been studied extensively in the literature \cite{gesbert2007adaptation, kiani2008optimal}. These optimization problems are well-known to be difficult because there is no known convex transformation available due to the inherit non-convexity of the SINR expression. However, a particular formulation which minimizes the total transmit power subject to the SINR constraints on single-antenna users over frequency flat channels is known to have an efficient global optimal solution by exploring semi-definite relaxation (SDR), second-order cone programming and uplink-downlink duality theory \cite{Gesbert2010, bengtsson1999optimal, dahrouj2010coordinated, wiesel2006linear}. There are also some works that study coordianted beamforming designs with distributed implementation in multicell network. For instance, in \cite{dahrouj2010coordinated}, coordinated beamforming with limited intercell information exchange is considered but the solution only applies to the sum-power minimization with SINR constraints. In \cite{huang2012distributed}, a distributed algorithm is designed to achieve the max-min rate fairness point on the Pareto boundary by means of uplink-downlink duality.

On the other hand, cooperative MIMO is a technique which has superior performance compared with the coordinated beamforming schemes at the expense of a huge backhaul loading requirement. There is much literature devoted to reducing the backhaul loading in cooperative MIMO schemes without sacrificing the merit of interference exploitation. For example, clustered multi-cell cooperation \cite{zhang2009networked,venkatesan2007coordinating,papadogiannis2008dynamic} has been proposed to reduce the backhaul loading by limiting the cooperation size. The clustering of BSs can be either static \cite{venkatesan2007coordinating} or dynamic \cite{papadogiannis2008dynamic}. While the intra-cluster interference can be effectively mitigated, the system capacity is still limited by the inter-cluster interference \cite{Gesbert2010}. Recently, distributed multicell processing schemes with data sharing are also developed in \cite{ng2008distributed,bjornson2010cooperative}. In \cite{ng2008distributed}, the distributed downlink beamforming design is recasted into a linear minimum mean-square-error (LMMSE) estimation problem.  \cite{bjornson2010cooperative} devised a distributed design with close-to-optimal performance using only local CSI. While these designs achieve relatively good performance gains, the huge backhaul cost due to data sharing between the BSs still remains.

Another example is called {\em opportunistic cooperation} which dynamically engages the system in different cooperation modes with various backhaul requirements, depending on the network channel conditions. For example, \cite{Gesbert2010} mentioned a few cooperation modes, such as interference coordination, full cooperation, rate-limited cooperation and relay-assisted cooperation.

In this paper, we consider the minimization of the backhaul cost associated with MIMO cooperation subject to the SINR constraints of all the mobiles in the downlink of multi-cell networks. Unlike conventional works on MIMO cooperation, we consider asymmetric cooperation between base stations so as to allow more flexibility to reduce the backhaul requirement. An example of asymmetric cooperation between base stations is illustrated in Figure 1, where BS-1 shares the data stream $s_{1}$ with BS-2 so as to cooperatively exploit the strong interference link to MS-1. BS-2 does not share the data stream in return because the interference link from BS-1 to MS-2 is weak. Directional MIMO cooperation has been considered in \cite{chowdhery2011cooperative}. A heuristic algorithm is proposed to dynamically select the directional cooperation links under a finite-capacity backhaul, subject to the evaluation of benefits and costs. However, the simple zero-forcing (ZF) transmit beamforming design assumed in \cite{chowdhery2011cooperative} in the directional cooperation cannot completely eliminate the undesired interference as in the full cooperation case. This is because the directional cooperation will result in data propagation which cannot be handled by a simple ZF scheme, and the true design is highly non-trivial in a directional cooperation topology. There are various technical challenges associated with the minimization of backhaul loading subject to SINR constraints in MIMO cooperation. They are elaborated as follows:
\begin{figure}[t]
\centering
\includegraphics[width=2.5in]{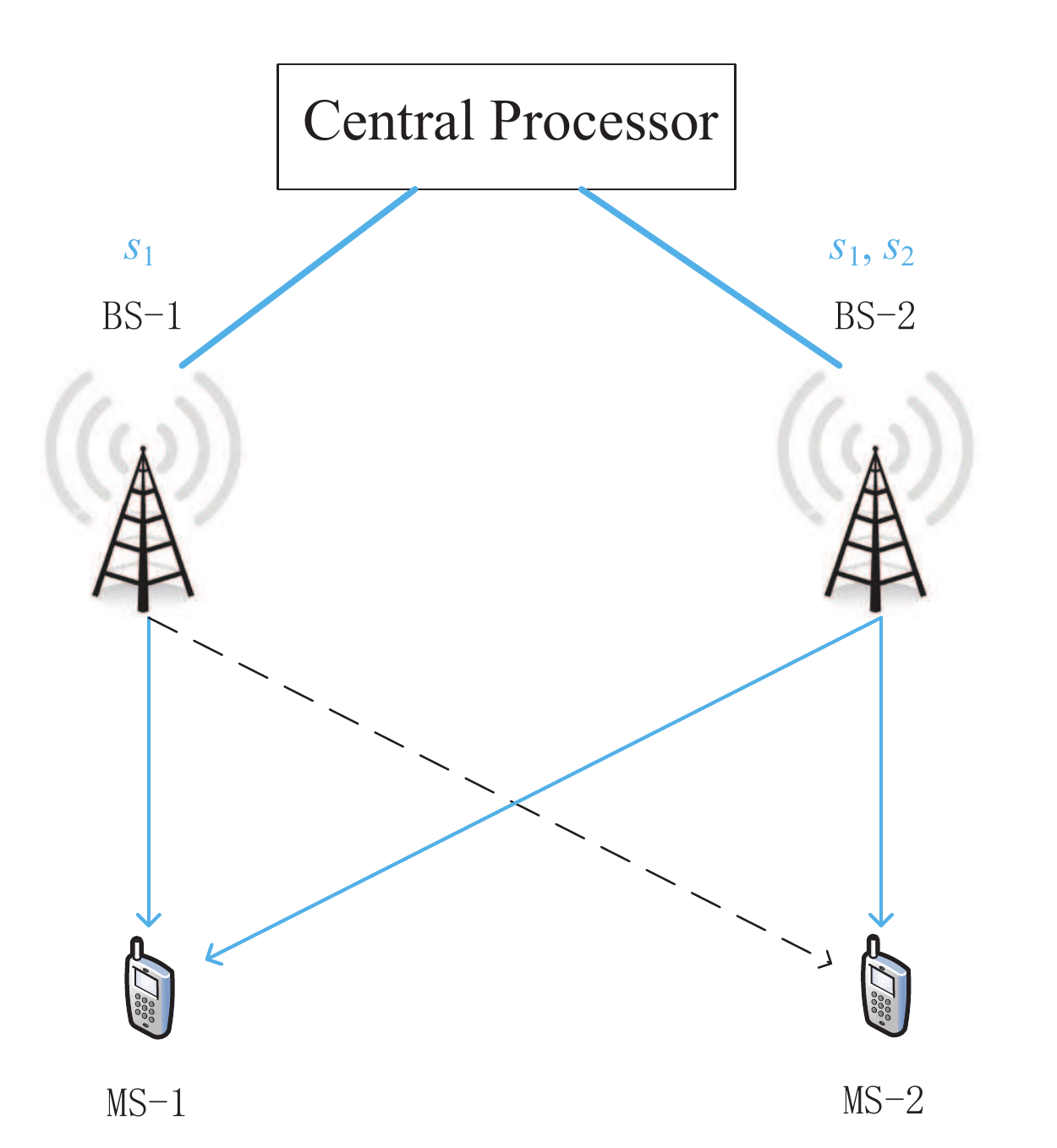}
\caption{An example of asymmetric directional cooperation. BS-1 shares its data stream with BS-2, while BS-2 does not share in return. } \label{fig:fig1}
\end{figure}

\begin{enumerate}
\item {\bf Appropriate Metric of Backhaul Cost:}
To quantify the backhaul cost, one obvious metric of backhaul loading is the average b/s/Hz in the backhaul to support the data streams to the mobiles. However, such a metric reveals too much detail, and the associated problem is highly combinatorial \cite{sanderovich2009uplink}. In this paper, we considers another metric, namely the number of {\em active directional cooperation links}. Since the number of active directional cooperation links is proportional to the backhaul loading (number of data streams) in the MIMO cellular network, this metric gives a first order measurement of the backhaul loading (in degrees of freedom) required to support the asymmetric MIMO cooperation without revealing too much unnecessary detail.
\item {\bf Combinatorial Optimization due to Backhaul Cost:}
The proposed backhaul metric involves counting the number of active cooperation links from the CP to the BSs, and hence, the associated optimization problem is combinatorial in nature. In fact, the proposed metric can be expressed as the mixed $l_{0}/l_{2}$ norm of the beamforming vector at each BS. Hence, the associated problem becomes an $l_{0}/l_{2}$ norm minimization problems, which is NP-hard in general.

\item {\bf Non-convexity due to SINR Constraints:}
Another difficulty of the optimization problem is the SINR constraints (requirement) of all the mobiles in the systems. These SINR constraints are non-convex due to the interference coupling between the users. As such, this makes the $l_{0}/l_{2}$ norm minimization problem different from the standard form of compressed sensing recovery \cite{donoho2006compressed},\cite{candes2008restricted}, and therefore, these standard solutions cannot be directly applied.
\end{enumerate}

In this paper, we tackle the above challenges of $l_{0}/l_{2}$ norm optimization under non-convex SINR constraints using the smooth approximation method \cite{chen2012smoothing, mohimani2009fast} and SDR \cite{bengtsson1999optimal, huang2010rank}.  Specifically, we show that the original $l_{0}/l_{2}$ norm minimization problem is asymptotically equivalent to a sequence of smooth minimization problems. We derive low complexity solutions by the SDR of the non-convex SINR constraints and show that the solution of the relaxed problem is always tight\footnote{In \cite{bengtsson1999optimal}, the authors established the strong duality results and the tightness of homogeneous SDR. In our case, the problem is inhomogeneous and we have extended the proof for strong duality under sufficient conditions.}  with probability 1. Convergence of the low complexity algorithm to the stationary point of the original non-smooth problem is guaranteed. It is worth noting that in a very recent paper \cite{zhaocoordinated}, the authors consider a similar problem scenario but with substantially different formulation and solution approach. A reweighted $l_{1}$ norm minimization method and a heuristic iterative link removal algorithm are proposed in \cite{zhaocoordinated} but these suboptimal schemes still suffer from a significant performance loss. The simulation result shows that our proposed algorithm can effectively reduce the backhaul loading required in the asymmetric MIMO cooperation compared with various baselines.

This paper is organized as follows. In Section II, we introduce the system model and formulate the combinatorial optimization problem, which minimizes the cooperation links subject to SINR constraints. In Section III, we use the smooth functions to approximate the $l_{0}/l_{2}$ norm and apply SDR to the non-convex quadratic constraints to obtain the approximation problem. In Section IV, a low complexity algorithm is proposed to solve the combinatorial  problem. Based on this, we show that the SDR is always tight with probability 1. Section V investigates the performance of the algorithm by simulation results. Section VI draws some concluding remarks.

\textit{Notation:} We adopt the notation of using boldface for vectors $\mathbf{a}$ (lower case) and matrices $\mathbf{A}$ (upper case). The transpose operator and the complex conjugate transpose operator are denoted by the symbols $(\cdot)^{T}$ and $(\cdot)^{H}$, respectively. $\textrm{Tr}(\cdot)$ is the trace of the square matrix argument. $\mathbf{I}$ and $\mathbf{0}$ denote, respectively, the identity matrix and the matrix with zero entries (their size is determined from context). $\mathbf{1}$ represents a vector of 1s and its size depends on the context. The kronecker product is denoted as $\bigotimes$. For any complex vector $\mathbf{x}$, we use $\|\mathbf{x}\|$, $\|\mathbf{x}\|_{0}$, $\|\mathbf{x}\|_{2}$ to represent the Euclidean norm, $l_{0}$ norm and $l_{2}$ norm of $\mathbf{x}$. The curled inequality symbol $\succeq$ (its strict form $\succ$ and reverse form $\prec$) is used to denote generalized inequality: $\mathbf{A}\succeq \mathbf{B}$ means that $\mathbf{A}-\mathbf{B}$ is a Hermitian positive semidefinite matrix ($\mathbf{A}\succ \mathbf{B}$ for positive definiteness and $\mathbf{A}\preceq \mathbf{B}$ for negative semidefiniteness). The $\dag$ symbol is used to denote pseudo-inverse: $\mathbf{A}^{\dag}$ means the pseudo-inverse of $\mathbf{A}$. $\mathcal{N}$ and $\mathcal{U}$ denotes Gaussian and uniform distribution respectively.

\section{System Model and Preliminaries}
\subsection{Asymmetric MIMO Cooperation}
\begin{figure}
    \centering
    \subfigure[]{
    \label{fig:subfig:a}
    \includegraphics[width=3.0in]{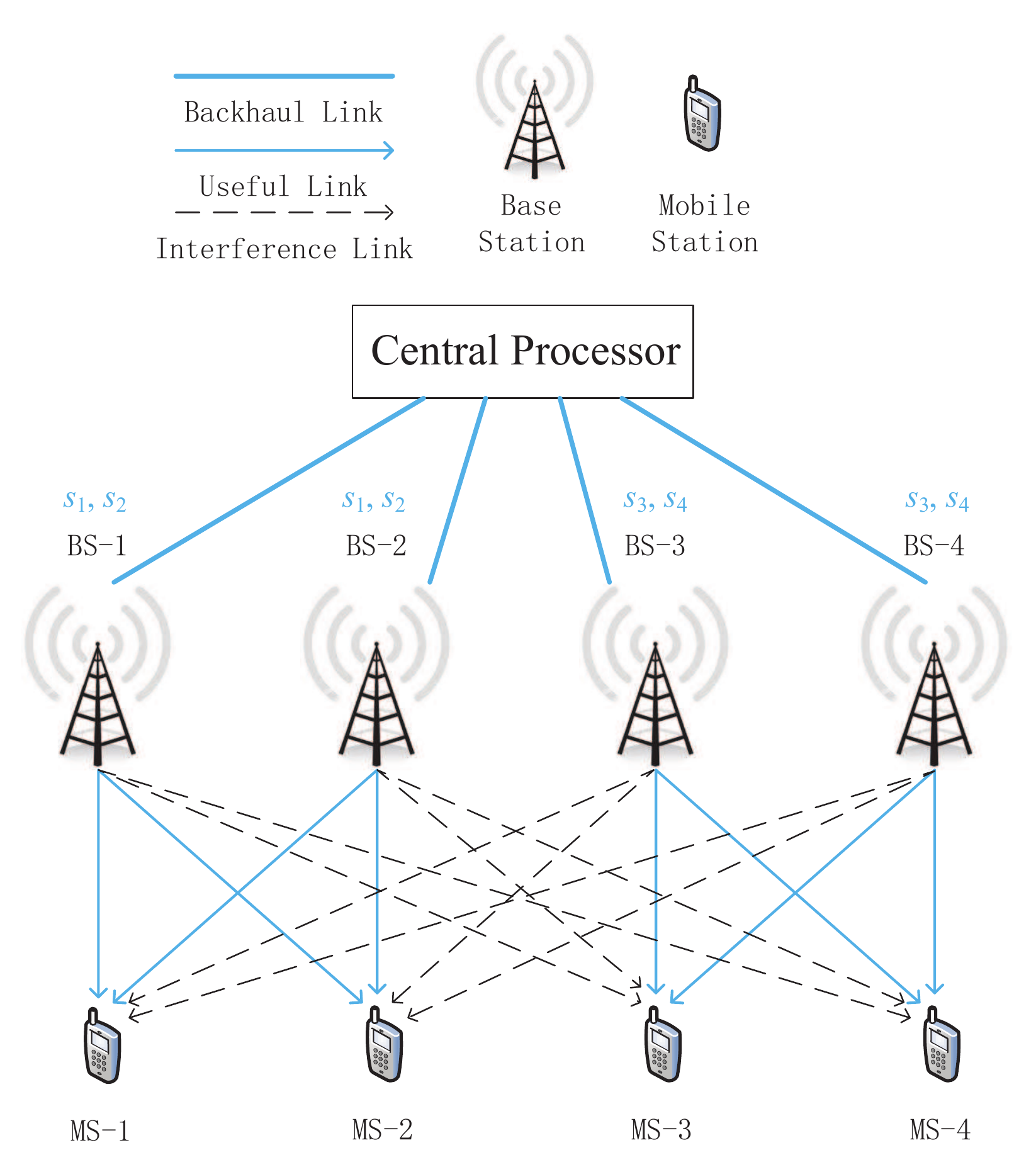}}

    \subfigure[]{
    \label{fig:subfig:b}
    \includegraphics[width=3.0in]{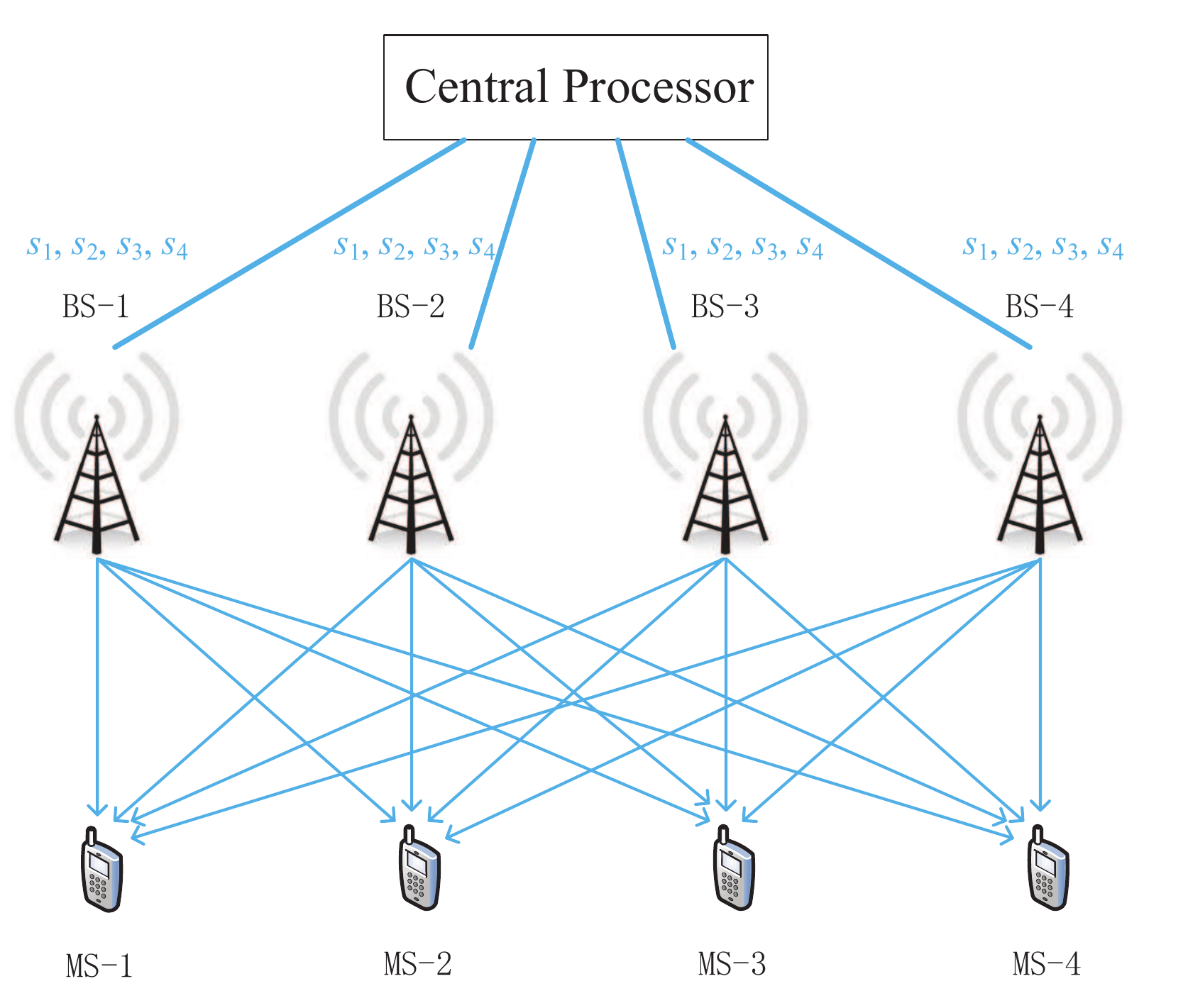}}

    \subfigure[]{
    \label{fig:subfig:c}
    \includegraphics[width=3.0in]{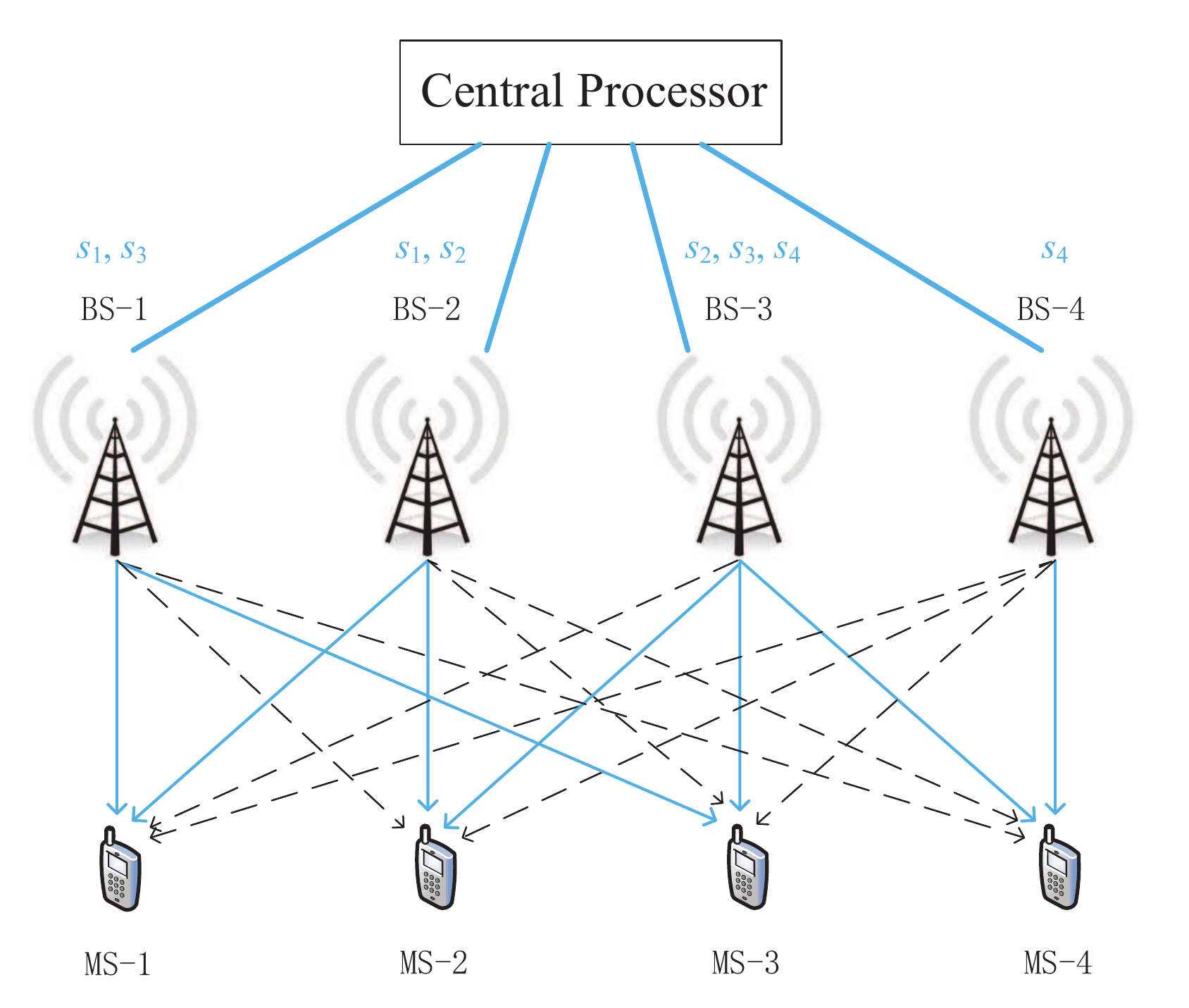}}
    \caption{Three multicell cooperation scenarios: (a) Cluster MIMO with two BSs forming a cooperation cluster. (b) Conventional full cooperation MIMO with each BS sharing information with all the other BSs. (c) Asymmetric MIMO topology with partial directional cooperation links activated. }
    \label{fig:subfig}
\end{figure}

Consider an MCP wireless system where several base stations jointly serve a number of single-antenna mobile stations (MSs) over a common frequency band, as shown in Figure \ref{fig:subfig}. Let the number of base stations be $N$, each equipped with $L$ transmit antennas, and let the total number of users in the system be $K$. We consider a centralized MCP backhaul model with a central processor (CP). All the user data and CSI are available at the CP for joint processing. Each BS is connected to the CP via backhaul links. The CSI is assumed to be perfect in this paper.

We consider {\em asymmetric MIMO cooperation} among the $N$ BSs in the MCP network to mitigate interference. Denote the set of BSs that have acquired the data signal for the $k$-th user as $\mathcal{Q}_{k}$, where $\mathcal{Q}_{k}\subseteq \{1,2,\cdots,N\}$. Specifically, cooperation is asymmetric, and the $\mathcal{Q}_{k}$ could be any subset of $\{1,2,\cdots,N\}$. For example, in Figure 2(c), the cooperation is asymmetric since $\mathcal{Q}_{1}=\{1,2\}$, $\mathcal{Q}_{2}=\{2,3\}$, $\mathcal{Q}_{3}=\{1,3\}$ and $\mathcal{Q}_{4}=\{3,4\}$. Therefore, the $n$-th BS that is  in $\mathcal{Q}_{k}$ can increase the received SINR of the $k$-th user by exploiting the link from the $n$-th BS to the $k$-th user, but it also incurs the backhaul cost of transmitting the data signal from the CP over the backhaul. Conventional full Cooperative-MIMO (Co-MIMO) and clustered Co-MIMO schemes are special cases of the above {\em asymmetric MIMO cooperation}, as illustrated in Figure \ref{fig:subfig:a} and \ref{fig:subfig:b}. In general, full symmetric MIMO cooperation is not always necessary, as previously illustrated in Figure \ref{fig:fig1}. Using the proposed asymmetric MIMO cooperation, flexible cooperation patterns can be engaged to satisfy the SINR requirements of users with minimum backhaul cost.

Let $s_{k} \in \mathbb{C}$ represents the information signal for the $k$-th user with unit energy. Let $\mathbf{w}_{n,k} \in \mathbb{C^{\textit{L}}}, \forall n=1,...,N$, be the beamforming vector used by the $n$-th BS to precode the information signal for the $k$-th user, and $\mathbf{w}_{k}=[ \mathbf{w}_{1,k}^{T}, \mathbf{w}_{2,k}^{T}, ..., \mathbf{w}_{n,k}^{T}]^{T} \in \mathbb{C^{\textit{LN}}}$ is the aggregate beamforming vector of the $N$ BSs for the $k$-th user. Note that $\mathbf{w}_{n,k}=\mathbf{0}$ if there is no data signal of the $k$-th user at the $n$-th BS. Let $\mathbf{h}_{n,k}\in \mathbb{C^{\textit{L}}}$ denote the complex channel fading between the $n$-th BS and the $k$-th user. The vector $\mathbf{h}_{k}=[\mathbf{h}_{1,k}^{T}, \mathbf{h}_{2,k}^{T}, ..., \mathbf{h}_{n,k}^{T}]^{T} \in \mathbb{C^{\textit{LN}}}$ is the aggregate channel fading from all the $N$ BSs to the $k$-th user, which is modeled as $\mathcal{CN}(0,I)$. The channel is assumed to be static over the transmission period. The received signal at the $k$-th user is given by
\begin{equation}
y_{k}=\underbrace{\mathbf{h}^{H}_{k}\mathbf{w}_{k}s_{k}}_{desired\,signal}
+\underbrace{\sum_{m\neq k}\mathbf{h}^{H}_{k}\mathbf{w}_{m}s_{m}}_{intercell\,interference}+n_{k}, \quad \forall k
\end{equation}

\noindent where $n_{k}$ is the complex Gaussian white noise with zero mean and variance  $\sigma^{2}_{k}$. The SINR at the $k$-th user is expressed as
\begin{equation}
\textrm{SINR}_{k}\triangleq\frac{|\mathbf{h}^{H}_{k}\mathbf{w}_{k}|^{2}}{\sum_{m\neq k}|\mathbf{h}^{H}_{k}\mathbf{w}_{m}|^{2}+\sigma^{2}_{k}}, \quad \forall k=1,...,K.
\end{equation}

\subsection{Conventional Optimal Beamforming Problem with Full Cooperation}
If there is no backhaul cost, all the BSs should participate in cooperative MIMO processing and they should mutually share all their data streams over the network. This corresponds to the following design problem that minimizes the total power consumption subject to the SINR requirements of each user:
\begin{align}
\label{OBP}
{\textrm{(OBP)}} \;
\mathop{\textrm{min}}_{\{\mathbf{w}_{k}\}^{K}_{k=1}} \quad & \sum_{k=1}^{K}\|\mathbf{w}_{k}\|_{2}^{2}   \\
\textrm{s.t.} \qquad & \frac{|\mathbf{h}^{H}_{k}\mathbf{w}_{k}|^{2}}{\sum_{m\neq k}|\mathbf{h}^{H}_{k}\mathbf{w}_{m}|^{2}+\sigma^{2}_{k}}\geqslant\gamma_{k}, \forall k. \nonumber
\end{align}

This optimal beamforming problem can be reformulated into a separable homogeneous Quadratically-Constrained-Quadratic-Program (QCQP), which is a non-convex problem. Furthermore, it was shown in \cite{luo2007approximation} that a separable homogeneous QCQP problem is NP-hard in general. Nonetheless, it is elegantly solved by \cite{bengtsson1999optimal} using the Perron-Frobenius theory for matrices with nonnegative entries. Generalized uplink-downlink duality using the Lagrangian theory is introduced in \cite{dahrouj2010coordinated} to efficiently solve the above (OBP). It is also worth noting that when the channel vectors are confined to the far-field, line-of-sight scenario (i.e., Vandermonde channel), the OBP can be reformulated as a convex problem \cite{Karipidis2007}.

However, in reality, the full sharing of user data is costly and highly impractical. It is possible and desirable to meet the same SINR constraints for each user, while requiring partial sharing of user data. This is a challenging combinatorial beamforming design problem, and it will be discussed in the following subsection $D$.
\subsection{Optimal Beamforming with Asymmetric Cooperation}
For a given asymmetric cooperation topology, which means all the $\mathcal{Q}_{k}$ are fixed in the MCP network with partially shared data signals, partial entries in the beamforming weights $\mathbf{w}_{k}$ should be forced to zero. Let $\mathbf{m}_{k} = [m_{1,k}\, m_{2,k} \, ...\, m_{N,k}]\in \mathbb{C}^{1\times N}$, where $m_{n,k}\in \{0,1\}$. Specifically, $m_{n,k}=0$ means that the $n$-th BS \emph{has} obtained the information signal for the $k$-th user while $m_{n,k}=1$ means the opposite. Thus, we have
\begin{equation}
\|\mathbf{m}_{k}\|_{0}\leq N-1
\end{equation}

\noindent since at least one BS is used to transmit signal to the $k$-th user. Let $\tilde{\mathbf{m}}_{k}=\mathbf{m}_{k}\bigotimes \mathbf{1}_{1\times L}$ and $\mathbf{M}_{k}=\mathrm{diag}(\tilde{\mathbf{m}}_{k})$. Therefore, we have
\begin{equation}
  \mathbf{M}_{k}\succeq0 \quad \textrm{and}\quad \textrm{rank}(\mathbf{M}_{k})\leq LN-L.
\end{equation}

Like in the OBP with full cooperation case, the design problem is to minimize the total power subject to the QoS requirements of each user, which gives the following optimization problem:
\begin{align}
\label{OBP-AC}
{\textrm{(OBP-AC)}} \quad
\mathop{\textrm{min}}_{\{\mathbf{w}_{k}\}^{K}_{k=1}} \quad & \sum_{k=1}^{K}\|\mathbf{w}_{k}\|_{2}^{2}   \\
\textrm{s.t.} \qquad & \frac{|\mathbf{h}^{H}_{k}\mathbf{w}_{k}|^{2}}{\sum_{m\neq k}|\mathbf{h}^{H}_{k}\mathbf{w}_{m}|^{2}+\sigma^{2}_{k}}\geqslant\gamma_{k}, \forall k \nonumber\\
& \mathbf{M}_{k}\mathbf{w}_{k}=\mathbf{0}, \forall k. \label{eq:obp-ac-const}
\end{align}

\noindent where the constraint (\ref{eq:obp-ac-const}) is to force the corresponding entries in $\mathbf{w}_{k}$ to be zero. Note that $\mathbf{M}_{k}$ are different matrices, which means different $\mathbf{w}_{k}$ may have different deactivated entries. Thus this optimization problem cannot be degenerated to (OBP) by simply omitting the deactivated entries and reducing the size of $\mathbf{w}_{k}$. This separable homogeneous QCQP problem with additional equality constraints (individual shaping constraints as termed in \cite{huang2010rank}) is still NP-hard. However, we manage to prove that the SDR is tight with probability 1, which will be discussed in detail in Section IV.B.
\subsection{Optimal Beamforming with Backhaul Cost}

We first define the backhaul cost associated with the asymmetric MIMO cooperation described in the subsection $A$. For a given asymmetric MIMO cooperation scheme $\{\mathbf{w}_{1},\mathbf{w}_{2},...,\mathbf{w}_{K}\}$, the backhaul cost is defined as the degree of freedom (number of data streams) over the backhauls from the CP to the $N$ BSs.

When $\mathbf{w}_{n,k}\neq\mathbf{0}$, it means that the data signal for the $k$-th user has to be shared from the CP to the $n$-th BS and this would consume backhaul for one unit using the cost metric we defined above. Correspondingly, when $\mathbf{w}_{n,k}=\mathbf{0}$, the user data for the $k$-th user will not be shared from the CP to the $n$-th BS, and hence, it will not consume the backhaul. This specific structure design on the beamforming vectors has recently drawn a lot of attention (see \cite{hong2013joint} for backhaul reduction and \cite{Mehanna2013} for antenna selection) and it is usally solved via the $l_{0}/l_{q}$ norm\footnote{Note any $l_{0}/l_{q}$ norm is a group sparse norm with $q\geq0$.} approach. In this paper, we consider the common setup with $q=2$ \cite{yuan2006model,hong2013joint}. Let $\tilde{\mathbf{w}}_{k}=[\|\mathbf{w}_{1,k}\|_{2},\|\mathbf{w}_{2,k}\|_{2},\cdots,\|\mathbf{w}_{N,k}\|_{2}]^{T}\in \mathbb{C}^{N}$. Then $\|\mathbf{w}_{k}\|_{0,2}=\|\tilde{\mathbf{w}}_{k}\|_{0}$. Since at least one BS should acquire the data signal $s_{k}$ from the CP to serve the $k$-th user, one of the backhaul usages of the $k$-th user should not be counted as additional backhaul cost. Thus, the backhaul \emph{cooperation} cost metric is given below:
\begin{equation}
C_{B}=\sum_{k=1}^{K}\|\mathbf{w}_{k}\|_{0,2}-K,
\end{equation}

\noindent where the term $K$ is to subtract the associated BSs from the cooperating backhaul cost. Specifically, we consider the following optimization problem:
\begin{align}
\label{P0}
{\textrm{(P0)}} \quad
\mathop{\textrm{min}}_{\{\mathbf{w}_{k}\}^{K}_{k=1}} \quad & \sum_{k=1}^{K}\|\mathbf{w}_{k}\|_{0,2}+\epsilon\sum_{k=1}^{K}\|\mathbf{w}_{k}\|_{2}^{2}   \\
\textrm{s.t.} \qquad & \frac{|\mathbf{h}^{H}_{k}\mathbf{w}_{k}|^{2}}{\sum_{m\neq k}|\mathbf{h}^{H}_{k}\mathbf{w}_{m}|^{2}+\sigma^{2}_{k}}\geqslant\gamma_{k}, \forall k.\nonumber
\end{align}

\noindent where $\epsilon>0$ is the power cost with respect to the total transmit power in the cellular network. This power cost term can be tuned to control the tradeoff between the backhaul cooperation cost and power consumption. An interesting relation between (OBP) and (P0) is that the $l_{2}$ norm used for the power minimization in (\ref{OBP}) is actually an early approach to approximate the sparest solution and is called the method of frames \cite{chen1998atomic}. Thus, the solution of (OBP) actually gives a rough estimate for the problem (P0).

In this paper, we are going to solve (P0) and the key challenges are due to the combinatorial $l_{0}/l_{2}$ norm minimization, which is NP-hard, and the non-convex homogeneous quadratic constraints. The non-convex SINR constraints turn (P0) into a non-standard $l_{0}/l_{2}$ norm minimization problem that is different from the traditional compressed sensing problem, where linear equality constraints are assumed. In \cite{Mehanna2013}, an $l_{1}/l_{\infty}$ norm\footnote{Note $l_{1}/l_{q}$ norm with $q>1$ is a group sparsity inducing norm \cite{bach2011optimization}.} squared norm approach is proposed to approximate the $l_0/l_{\infty}$ norm and the problem is solved using SDP transformations. However, this approach cannot be directly extended to the more general multi-antenna MCP scenario we consider in this paper, because the approach in \cite{Mehanna2013} relies on coupling across all the beamforming vectors due to the group norm. Thus, so far, there is no reliable and general method to deal with $l_{0}/l_{2}$ (or any other $l_{0}/l_{q}$) norm minimization with nonlinear and non-convex constraints. In this paper, we shall first introduce an approximation problem of (P0) and then deploy the SDR technique to solve the approximation problem. Later we prove that the approximation problem and the original problem (P0) are asymptotically equivalent.

\section{Problem Approximation and Transformation}
\subsection{Transformation 1 - Approximating the $l_{0}$ Norm with Smooth Functions}
The problem of using the $l_{0}/l_{2}$ norm for a combinatorial search for its minimization is due to the fact that the $l_{0}$ norm is a non-convex and non-smooth function. \cite{chen2012smoothing} and \cite{mohimani2009fast} state an idea to approximate this non-smooth $l_{0}$ by a suitable continuously differentiable function, and minimize it by means of calculus (e.g., steepest descent method). The advantage of smoothing methods is that we solve optimization problems with continuously differentiable functions for which there are rich theory and powerful solution methods \cite{nocedal1999numerical}. A local minimizer or stationary point of the original non-smooth problem can be guaranteed to be found by updating the smooth parameter \cite{chen2012smoothing}. The efficiency of smoothing methods will depend on the smooth approximation function, the solution method for the smooth optimization problem and the updating scheme for the smoothing parameter, which is $\theta$ in our case.

In this paper, we consider the following family of smooth functions with the smoothing parameter $\theta$ \cite{mohimani2009fast}:
\begin{equation}
f_{\theta}(s)\triangleq \textrm{exp}(\frac{-s^{2}}{2\theta^{2}}).
\end{equation}

\noindent It is easily observed that
\begin{equation}
\mathop{\textrm{lim}}_{\theta \rightarrow 0}f_{\theta}(s)=\Big{\{}
\begin{array}{clcr}
1, & \textrm{if} \,s=0\\0, & \textrm{if}\, s\neq0.
\end{array}
\end{equation}



\noindent Then, by defining
\begin{equation}
\hat{F}_{\theta}(\mathbf{w}_{k})= \sum^{N}_{i=1}f_{\theta}(\|\mathbf{w}_{i,k}\|_{2}),
\end{equation}

\noindent and
\begin{equation}
F_{\theta}(\mathbf{w}_{1},...,\mathbf{w}_{K})= \sum^{K}_{k=1}\hat{F}_{\theta}(\mathbf{w}_{k}),
\end{equation}

\noindent it is readily verified that
\begin{equation}
\label{l0 norm approx}
\sum_{k=1}^{K}\|\mathbf{w}_{k}\|_{0,2}\approx KN - F_{\theta}(\mathbf{w}_{1},...,\mathbf{w}_{K})
\end{equation}
for small values of $\theta$, and the approximation is tight when $\theta \rightarrow 0$. Consequently, the minimum $l_{0}/l_{2}$ norm solution can be found by maximizing $F_{\theta}(\mathbf{w}_{1},...,\mathbf{w}_{K})$ (subject to the SINR constraints) for a sufficiently small value of $\theta$. Note that the value of $\theta$ determines the smoothness of the function $F_{\theta}$. The larger the value of $\theta$, the smoother $F_{\theta}$, but the worse the approximation to the $l_{0}/l_{2}$ norm. On the other hand, the smaller the value of $\theta$, the closer the behavior of $F_{\theta}$ to the $l_{0}/l_{2}$ norm. This smooth maximization problem can be solved using a projected gradient algorithm.  \cite{mohimani2009fast} and \cite{chen2012smoothing} propose an idea to use a decreasing sequence for the smooth parameter $\theta$ to gradually approach the actual maximum for smaller values of $\theta$.

Replacing the $l_{0}/l_{2}$ norm objective with $F_{\theta}$, we have
\begin{align}
\label{P1}
{(\textrm{P1}(\theta))} \quad
\mathop{\textrm{max}}_{\{\mathbf{w}_{k}\}^{K}_{k=1}} \quad & F_{\theta}(\mathbf{w}_{1},...,\mathbf{w}_{K})-\epsilon\sum_{k=1}^{K}\|\mathbf{w}_{k}\|_{2}^{2}   \\
\textrm{s.t.} \quad & \frac{|\mathbf{h}^{H}_{k}\mathbf{w}_{k}|^{2}}{\sum_{m\neq k}|\mathbf{h}^{H}_{k}\mathbf{w}_{m}|^{2}+\sigma^{2}_{k}}\geqslant\gamma_{k}, \forall k.\nonumber
\end{align}

Let $\mathbf{x}=[\mathbf{w}_{1}^{T}\, \mathbf{w}_{2}^{T}\, ...\, \mathbf{w}_{K}^{T}]^{T} \in \mathbb{C}^{KLN}$ and $\mathcal{W}$ (at the top of the next page) denote the feasible set of problems (P1$(\theta)$) and (P0). Let $p(\mathbf{x})=\|\mathbf{x}\|_{0,2}+\epsilon\|\mathbf{x}\|_{2}^{2}=\sum_{k=1}^{K}\|\mathbf{w}_{k}\|_{0,2}+\epsilon\sum_{k=1}^{K}\|\mathbf{w}_{k}\|_{2}^{2}$, then (P0) can be rewritten as
\begin{figure*}[t]
\begin{equation}
\mathcal{W}=\Big{\{}\mathbf{x}=[\mathbf{w}_{1}^{T}\, \mathbf{w}_{2}^{T}\, ...\, \mathbf{w}_{K}^{T}]^{T}\Big{|}\textrm{SINR}_{k}=\frac{|\mathbf{h}^{H}_{k}\mathbf{w}_{k}|^{2}}{\sum_{m\neq k}|\mathbf{h}^{H}_{k}\mathbf{w}_{m}|^{2}+\sigma^{2}_{k}}\geqslant\gamma_{k}, \quad \forall k=1,...,K\Big{\}}
\end{equation}
\hrulefill
\end{figure*}
\begin{equation}
\mathop{\textrm{min}}_{\mathbf{x}\in \mathcal{W}} \quad p(\mathbf{x}).
\end{equation}

We assume (P0) is always feasible and let $P^{*}$ be the set of such optimal solutions and  $g(\mathbf{x},\theta)=-F_{\theta}(\mathbf{x})+\epsilon\|\mathbf{x}\|_{2}^{2} $, then (P1$(\theta)$) is the same as
\begin{equation}
\mathop{\textrm{min}}_{\mathbf{x}\in \mathcal{W}} \quad g(\mathbf{x},\theta).
\end{equation}

\textit{Assumption 1.}  There exists a finite set $S^{*} \subset C^{KLN}$ having the property that, for any $\theta \in \Theta\subseteq R$, a point $\mathbf{x}(\theta) \in S^{*}$ exists such that
\begin{equation}
\mathbf{x}(\theta) \in \textrm{arg}\ \mathop{\textrm{min}}_{\mathbf{x} \in \mathcal{W}}g(\mathbf{x},\theta).
\end{equation}

By Assumption 1, we assume that the feasible set $\mathcal{W}$ is always nonempty and there exists an optimal solution for problem (P1$(\theta)$) for any $\theta$ value.

\textit{Theorem 1 (Asymptotic Equivalence between (P1$(\theta$)) and (P0)).} (Theorem 3.2.1, \cite{rinaldimathematical}) Let $\{\theta^{j}\}\subset \Theta$ be an infinite sequence such that
\begin{equation}
\label{thm1}
\mathop{\textrm{lim}}_{j\rightarrow \infty}g(\mathbf{x},\theta^{j})=p(\mathbf{x}), \quad \forall \mathbf{x} \in \mathcal{W}.
\end{equation}

Under Assumption 1, there exists a finite index $\bar{j}$ such that, for any $j\geq \bar{j}$, problem (P1$(\theta)$), with $\theta = \theta^{j}$, has a solution $x^{j}$ that also solves the original problem (P0).


\textit{Definition 1: (Stationary point of non-smooth problem (P0))}: Stationary point of the non-smooth problem (P0) is defined as the limit of the stationary point of (P1$(\theta)$) as $\theta \rightarrow 0$.

Before introducing the algorithm to solve (P1$(\theta)$), note that the SINR quadratic constraints in (\ref{P1}) are non-convex, and hence, the problem (P1$(\theta)$) is difficult to solve. We shall apply SDR technique to relax the non-convex constraints.

\subsection{Transformation 2 - Semidefinite Relaxation of the Non-convex SINR Constraints in (P1$(\theta)$)}
We define
\begin{equation}
\label{eq:Wk=wkwkH}
\mathbf{W}_{k}\triangleq \mathbf{w}_{k}\mathbf{w}_{k}^H \quad \textrm{and}\quad
\mathbf{H}_{k}\triangleq \mathbf{h}_{k}\mathbf{h}_{k}^H, \quad \forall k=1,...,K
\end{equation}

Then (P1$(\theta)$) can be written as:
\begin{align}
\mathop{\textrm{max}}_{\{\mathbf{w}_{k}, \mathbf{W}_{k}\}^{K}_{k=1}} \; & F_{\theta}(\mathbf{w}_{1},...,\mathbf{w}_{K})-\epsilon\sum_{k=1}^{K}\|\mathbf{w}_{k}\|_{2}^{2}    \\
\textrm{s.t.} \qquad & \mathrm{Tr}(\mathbf{H}_{k}\mathbf{W}_{k})-\gamma_{k}\mathop{\sum}_{m\neq l}\mathrm{Tr}(\mathbf{H}_{k}\mathbf{W}_{m})-\gamma_{k}\sigma_{k}^{2} \geq 0 \nonumber \\ &
\mathbf{W}_{k}= \mathbf{w}_{k}\mathbf{w}_{k}^H, \forall k.\nonumber
\end{align}

This reformulated problem still has a non-convex constraint $\mathbf{W}_{k}= \mathbf{w}_{k}\mathbf{w}_{k}^H$. Now observe the following equivalence:
\begin{align}
\label{rank1_1}
\mathbf{W}_{k} =\mathbf{w}_{k}\mathbf{w}_{k}^H &\Leftrightarrow \mathbf{W}_{k}\succeq \mathbf{w}_{k}\mathbf{w}_{k}^H, \mathrm{rank}(\mathbf{W}_{k})\leq 1.\nonumber\\&\Leftrightarrow \mathbf{W}_{k}\succeq \mathbf{0}, \mathrm{rank}(\mathbf{W}_{k})\leq 1.
\end{align}

The constraint $\mathbf{W}_{k}\succeq \mathbf{w}_{k}\mathbf{w}_{k}^H$ is convex and can be formulated as a Schur complement \cite{boyd2004convex}:
\begin{equation}
\label{eq:LMI_PSD}
\left[\begin{array}{cccc}{\mathbf{W}_{k}}&{\mathbf{w}_{k}}\\{\mathbf{w}_{k}^H}&{1}\end{array}\right]\succeq\mathbf{0}.
\quad
\end{equation}

In SDR approximation, the rank constraint on $\mathbf{W}_{k}$ is dropped to obtain a relaxed problem:
\\ \indent (P2$(\theta)$)
\begin{align}\label{eq:p2theta}
\mathop{\textrm{max}}_{\{\mathbf{w}_{k}, \mathbf{W}_{k}\}^{K}_{k=1}} \quad & F_{\theta}(\mathbf{w}_{1},...,\mathbf{w}_{K})-\epsilon\sum_{k=1}^{K}\|\mathbf{w}_{k}\|_{2}^{2}    \\
\textrm{s.t.} \qquad & \mathrm{Tr}(\mathbf{H}_{k}\mathbf{W}_{k})-\gamma_{k}\mathop{\sum}_{m\neq k}\mathrm{Tr}(\mathbf{H}_{k}\mathbf{W}_{m}) -\gamma_{k}\sigma_{k}^{2} \geq 0 \nonumber\\ &
\left[\begin{array}{cccc}{\mathbf{W}_{k}}&{\mathbf{w}_{k}}\\{\mathbf{w}_{k}^H}&{1}\end{array}\right]\succeq\mathbf{0}, \quad\forall k.\nonumber
\end{align}

Thus, (P2$(\theta)$) is a smooth optimization problem. As mentioned above, an optimization problem with continuously differentiable functions has prosperous theoretical results and powerful methods, like the Frank-Wolfe algorithm \cite{rinaldimathematical} and the proximal gradient method\cite{beck2009fast}. In this paper, we use a projected gradient algorithm to solve (P2$(\theta)$).

\section{Low Complexity Solution}
\begin{algorithm*}
\caption{Proposed Algorithm}
\begin{itemize}
\item Step 1 \emph{Initialization}:
\begin{enumerate}
\item Let $\{\mathbf{w}_{k}^{0}\}_{k=1}^{K}$ be the $l_{2}$ norm solution, i.e., the solution of the optimization problem (OBP-SDP).

\item Set the $\theta$ decreasing factor $0<\eta<1$, $\tau>0$ and the perturbation factor $\varrho$.
\end{enumerate}
\item Step 2 \emph{Projection gradient loop}: For {$j=1,...,J$}
\begin{enumerate}
\item Let $\theta=\theta^{j}$.
\item \textit{Projection gradient step}: Maximize the objective $F=(F_{\theta}(\mathbf{w}_{1},...,\mathbf{w}_{K})-\epsilon\sum_{k=1}^{K}\|\mathbf{w}_{k}\|_{2}^{2})$ on the feasible set using the steepest ascent method followed by projection onto the feasible set.
    \begin{itemize}
        \item Find the gradient $\mathbf{\delta}_{k}=\nabla_{\mathbf{w}_{k}}F\in \mathbb{C}^{LN}$.
        \item Let the perturbation be $e(\mathbf{w}_{k}^{j-1},\mu^{j})=\mu^{j}\varsigma\mathbf{w}_{k}^{j-1}$, where $\varsigma\sim \mathcal{U}[-\varrho,\varrho]$.\footnotemark
        \item Let $\bar{\mathbf{w}}_{k}= \mathbf{w}_{k}^{j-1}+\mu^{j}\mathbf{\delta}_{k}-e(\mathbf{w}_{k}^{j-1},\mu^{j})$, $\forall k$

        \item Project $\bar{\mathbf{w}}_{k}$ back onto the feasible set to get $\mathbf{w}_{k}^{j}$, i.e., equivalent to solving the projection problem (AP) (\ref{eq:AP}).
    \end{itemize}
    \item \textit{$\theta$ updating step}: if $\|\sum_{k=1}^{K}\mathbf{\delta}_{k}(\mathbf{w}_{k}^{j}-\mathbf{w}_{k}^{j-1})\|\geq\tau\theta^{j-1}$, then set $\theta^{j}=\theta^{j-1}$; otherwise, choose $\theta^{j}=\eta\theta^{j-1}$.

\end{enumerate}
\item Step 3 \emph{Find the minimum power solution}: Obtain the positions of the zero entries in $\mathbf{w}_{k}^{J}$ to form $\mathbf{m}_{k}, \forall k$.
    \begin{enumerate}
        \item Use $\{\mathbf{m}_{k}\}_{k=1}^{K}$ to solve an (OBP-AC-SDP) and obtain the solution $\{\breve{\mathbf{w}}_{k}^{J}\}_{k=1}^{K}$.
    \end{enumerate}
\item The final solution is given by $\{\breve{\mathbf{w}}_{k}^{J}\}_{k=1}^{K}$.
\end{itemize}
\end{algorithm*}

\begin{figure}[ht]
\centering
\includegraphics[width=3.5in]{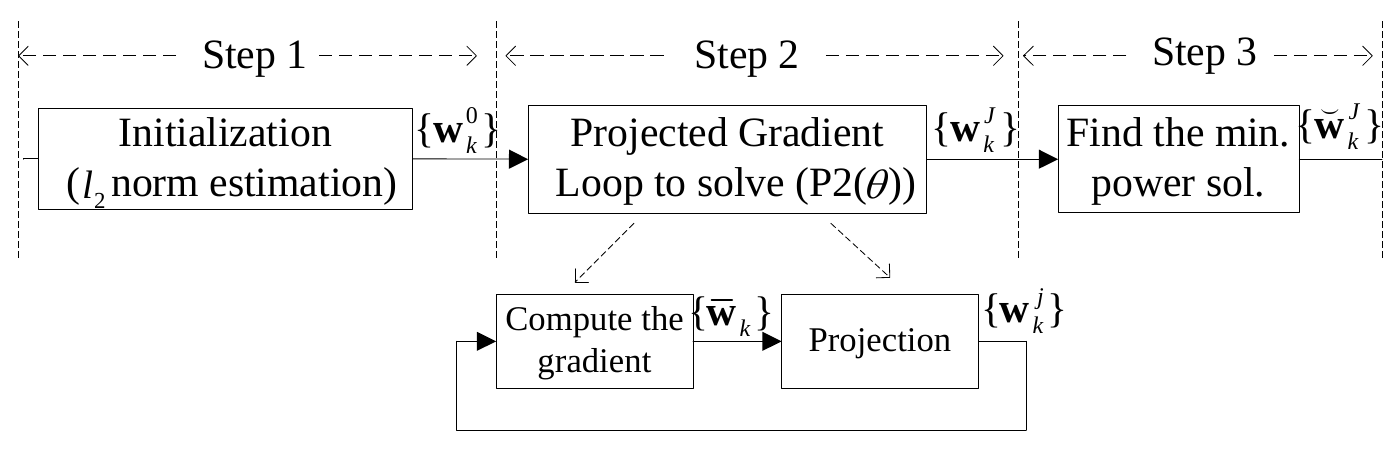}
\caption{Block diagram for Algorithm 1.} \label{fig:fig_block_diag1}
\end{figure}

In the following, we shall introduce Algorithm 1 to solve (P2$(\theta)$), which is shown below. Step 1 is to solve an $l_{2}$ norm estimation of (P0) as an initial point. Step 2 is the projected gradient loop, which is used to find the solution of the smooth optimization (P2$(\theta)$) for a decreasing sequence of $\theta$. Finally, based on the obtained backhaul cooperation solution, Step 3 is to find the minimum power consumption solution. An illustrative block diagram of Algorithm 1 is summarized in Figure \ref{fig:fig_block_diag1}.

The $l_{2}$ norm estimation of (P0) in Step 1 is actually the problem (OBP), which serves an rough estimate for the initial point of Algorithm 1. The projected gradient loop in Step 2 is equivalent to repeatedly solve the projection problem (AP), which is shown as follows:
\begin{align}\label{eq:AP}
{\textrm{(AP)}} \quad \mathop{\textrm{min}}_{\{\mathbf{w}_{k}, \mathbf{W}_{k}\}^{K}_{k=1}} \quad & \sum_{k=1}^{K}\|\mathbf{w}_{k}-\bar{\mathbf{w}}_{k}\|_{2}^{2}   \\
\textrm{s.t.} \quad & \textrm{Constraints in}\, (\ref{eq:p2theta}),\nonumber
\end{align}

\noindent where $\{\bar{\mathbf{w}}_{k}\}_{k=1}^{K}$ coming from the gradient computation step as illustrated in Figure \ref{fig:fig_block_diag1}, are the vectors to be projected. To obtain the minimum power solution, an (OBP-AC) is solved in Step 3. Note that (OBP) and (OBP-AC) are not convex due to the SINR constraints. On the other hand, we observe that the projection problem (AP) can be transformed into a pure SDP form, which is more amenable for rank characterization and analysis. We shall first perform the subproblem transformations next.

\subsection{Transformations of Subproblems in Algorithm 1}
\subsubsection{Semidefinite Relaxation in (OBP) and (OBP-AC)}
By using the relation in (\ref{eq:Wk=wkwkH}) and (\ref{rank1_1}) to drop the rank constraint, we can relax the non-convex (OBP) into the following form:\\
\indent (OBP-SDP)
\begin{align}\label{eq:OBP-SDP}
\mathop{\textrm{min}}_{\{\mathbf{W}_{k}\}^{K}_{k=1}} \quad & \sum_{k=1}^{K}\mathrm{Tr}(\mathbf{W}_{k})  \\
\textrm{s.t.} \qquad & \mathrm{Tr}(\mathbf{H}_{k}\mathbf{W}_{k})-\gamma_{k}\mathop{\sum}_{m\neq k}\mathrm{Tr}(\mathbf{H}_{k}\mathbf{W}_{m})-\gamma_{k}\sigma_{k}^{2} \geq 0 \nonumber\\ &
\mathbf{W}_{k}\succeq \mathbf{0}, \quad\forall k\nonumber
\end{align}

\noindent Similarly, (OBP-AC) can be relaxed into the following form:
\begin{align}
{\textrm{(OBP-AC-SDP)}} \quad \mathop{\textrm{min}}_{\{\mathbf{W}_{k}\}^{K}_{k=1}} \quad & \sum_{k=1}^{K}\mathrm{Tr}(\mathbf{W}_{k})  \\
\textrm{s.t.} \qquad & \textrm{Constraints in}\, (\ref{eq:OBP-SDP}). \nonumber\\ &
\mathrm{Tr}(\mathbf{M}_{k}\mathbf{W}_{k})=0,\quad\forall k\nonumber
\end{align}

\subsubsection{Transformation of (AP) into SDP Form}

Note that the objective of (AP) is inhomogeneous quadratic, which can be rewritten as
\begin{equation}
\|\mathbf{w}_{k}-\bar{\mathbf{w}}_{k}\|_{2}^{2} =[1 \quad \mathbf{w}_{k}^{H}]\left[\begin{array}{clcr}{\bar{\mathbf{w}}_{k}^{H}\bar{\mathbf{w}}_{k}}&{-\bar{\mathbf{w}}_{k}^{H}}
  \\{-\bar{\mathbf{w}_{k}}}&{\mathbf{I}}
  \end{array}\right]
  \left[\begin{array}{clcr}{1}\\{\mathbf{w}}_{k}\end{array}\right].
\end{equation}

Let
\begin{equation}
\label{inhmg:eq1}
\mathbf{I}_{00}=\left[\begin{array}{clcr}{1}&{\mathbf{0}^{H}}\\{\mathbf{0}}&{\mathbf{O}}\end{array}\right],
\quad
\end{equation}

\noindent where all entries of $\mathbf{O}\in \mathbb{S}^{N}$ and $\mathbf{0}\in \mathbb{C}^{N}$ are 0. Thus $\mathbf{I}_{00}\in \mathbb{S}^{N+1}$. Let $\mathbf{I}\in \mathbb{S}^{N}$,
\begin{equation}
\label{inhmg:eq2}
\mathbf{A}_{k}=\left[\begin{array}{clcr}{\bar{\mathbf{w}}_{k}^{H}\bar{\mathbf{w}}_{k}}&{-\bar{\mathbf{w}}_{k}^{H}}
  \\{-\bar{\mathbf{w}_{k}}}&{\mathbf{I}}\end{array}\right] \; \textrm{and} \;
\tilde{\mathbf{H}}_{k}=\left[\begin{array}{clcr}{0}&{\mathbf{0}^{H}}\\{\mathbf{0}}&{\mathbf{H}}_{k}\end{array}\right].
\end{equation}

\noindent Thus, $\mathbf{A}_{k},\mathbf{H}_{k}\in \mathbb{S}^{N+1}$. Let $
\tilde{\mathbf{w}}_{k}=[1 \quad {\mathbf{w}}_{k}^{T}]^{T}$ and
$\mathbf{\tilde{W}}_{k} =\mathbf{\tilde{w}}_{k}\mathbf{\tilde{w}}_{k}^H.$

Using the relation in (\ref{rank1_1}) to drop the rank constraint, the inhomogeneous optimization problem (AP) can be transformed into SDP form:\\
\indent (AP-SDP)
\begin{align}
\mathop{\textrm{min}}_{\{\tilde{\mathbf{W}}_{k}\}^{K}_{k=1}} \quad & \sum_{k=1}^{K}\mathrm{Tr}(\mathbf{A}_{k}\tilde{\mathbf{W}}_{k})  \label{a2-sdp}\\
\textrm{s.t.} \qquad & \mathrm{Tr}(\tilde{\mathbf{H}}_{k}\tilde{\mathbf{W}}_{k})-\gamma_{k}\mathop{\sum}_{m\neq k}\mathrm{Tr}(\tilde{\mathbf{H}}_{k}\tilde{\mathbf{W}}_{m})-\gamma_{k}\sigma_{k}^{2} \geq 0  \nonumber\\ &
\mathbf{Tr}(\mathbf{I}_{00}\tilde{\mathbf{W}}_{k})=1,\quad\forall k \label{a2-sdp-cons}\\
& \tilde{\mathbf{W}}_{k}\succeq \mathbf{0}, \quad\forall k\nonumber
\end{align}

Note that (AP) and (AP-SDP) are equivalent before dropping the rank constraint. As such, (AP) is replaced by solving (AP-SDP) in Algorithm 1. In the following subsection $B$, we will show that the rank relaxation for (AP-SDP) is always tight with probability 1.

\footnotetext{Adding this small perturbation is to ensure that the output of the SDR solution is rank 1 with probability 1, which will be shown in the proof of Theorem 3.}

\begin{figure}[ht]
\centering
\includegraphics[width=3.0in]{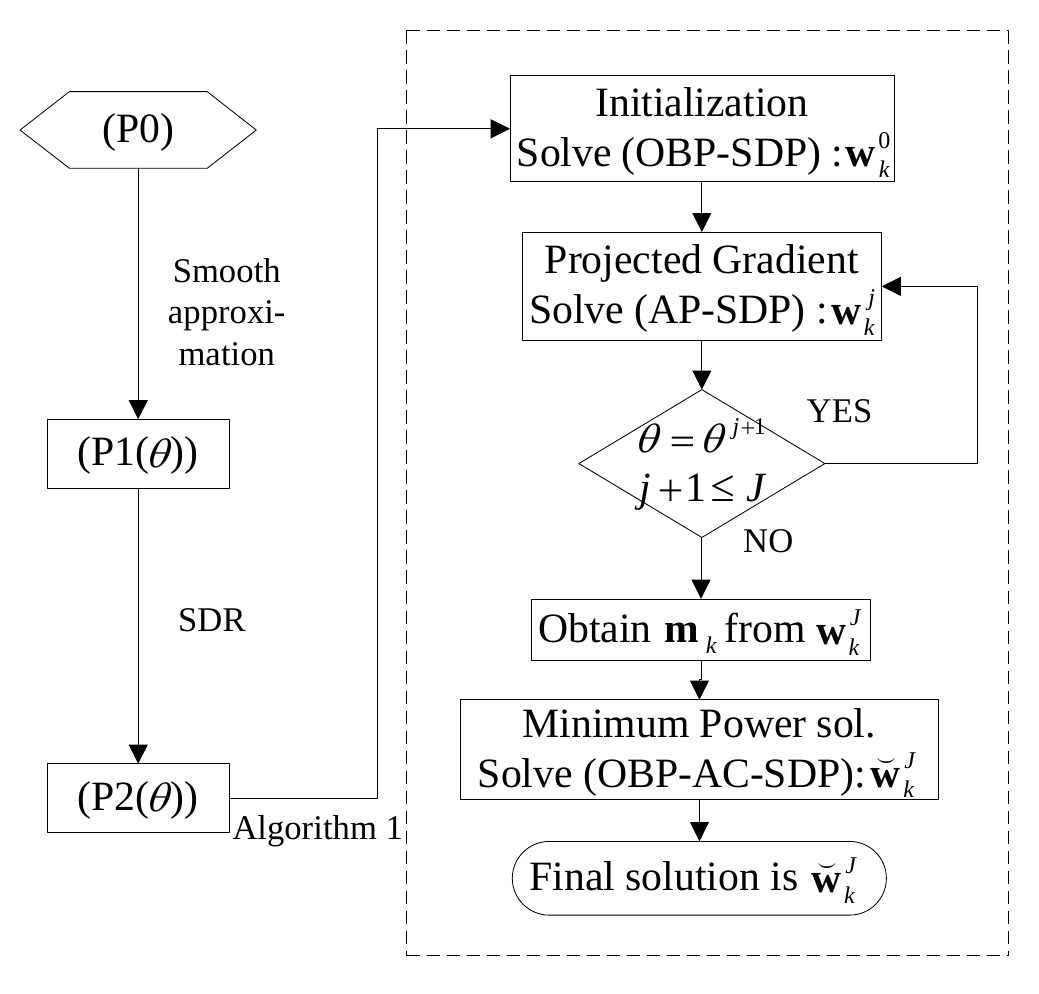}
\caption{Overall solution structure for solving (P0).} \label{fig:fig_block_diag2}
\end{figure}

The overall solution structure to solve (P0) is summarized in Figure \ref{fig:fig_block_diag2}.


\subsection{Discussion of the Optimality of Algorithm 1}
(OBP-SDP) is a power minimization problem subject to SINR requirements that is known to have rank 1 optimal solutions by using uplink-downlink duality and Perron-Frobenius theory for matrices with nonnegative entries \cite{bengtsson1999optimal}. For completeness, we provide a simple proof using the KKT conditions.

\textit{Theorem 2 (Strong duality\footnote{Strong duality means that all the optimal solutions are rank 1, i.e., the SDR relaxation is tight.} of (OBP-SDP))}: Suppose the optimal solution for (OBP-SDP) is $\{\mathbf{W}_{k}^{*}\}_{k=1}^{K}$, then $\textrm{rank}(\mathbf{W}_{k}^{*})=1, \forall k=1,...,K$.

\textit{Proof:}
The lagrangian dual problem (OBP-SDP) is shown below:\\
\indent (OBP-SDP-Dual)
\begin{align}
\label{A1-SDP-Dual}
\mathop{\textrm{max}}_{\{\lambda_{k}\}^{K}_{k=1}} \quad & \sum_{k=1}^{K}\lambda_{k}\gamma_{k}\sigma_{k}^{2} \nonumber\\
\textrm{s.t.} \qquad & \mathbf{Z}_{k}=\mathbf{I}-\lambda_{k}\mathbf{H}_{k}+\sum_{m\neq k}\lambda_{m}\gamma_{m}\mathbf{H}_{m}\succeq0,\quad \forall k \\
& \lambda_{k} \geq0, \quad\forall k\nonumber
\end{align}

$\lambda_{k}$ and $\mathbf{Z}_{k}$ are lagrangian multipliers for the SINR inequality and the PSD constraints, respectively. We know $\mathbf{H}_{k}=\mathbf{h}_{k}\mathbf{h}_{k}^{H}$ are all positive semidefinite and rank 1. Thus, $\mathbf{I}+\sum_{m\neq k}\lambda_{m}\gamma_{m}\mathbf{H}_{m}$ is obviously a full rank matrix. Therefore, $\mathrm{rank}(\mathbf{Z}_{k})\geq N-1$. By the KKT condition, we know $\mathbf{Z}_{k}\mathbf{W}_{k}^{*}=\mathbf{0}$. Because $\mathbf{W}_{k}^{*}\succeq0$, then $\mathrm{rank}(\mathbf{W}_{k}^{*})\leq 1$. $\Box$

The inhomogeneous objective results in an increase of the variable size by 1 and imposes an extra $K$ equality constraints to be homogenized into (AP-SDP). However, this SDR may not be tight due to the extra constraints, as we know the tightness of the SDR is strongly related to the number of constraints\cite{huang2010rank}. But in the following, we show that the SDR is tight with probability 1 in this particular problem.

\textit{Theorem 3 (Strong duality of (AP-SDP))}: Suppose the optimal solution for (AP-SDP) is $\{\tilde{\mathbf{W}}_{k}^{*}\}_{k=1}^{K}$, then $\textrm{rank}(\tilde{\mathbf{W}}_{k}^{*})=1, \forall k=1,...,K$ under sufficient condition that $\mathbf{H}_{k}\bar{\mathbf{w}}_{k}\neq\mathbf{0}$ (i.e., the desired beamforming vector $\bar{\mathbf{w}}_{k}$ is not orthogonal to the channel $\mathbf{H}_{k}$). This condition is always satisfied in Algorithm 1 with probability 1, i.e., (AP-SDP) has strong duality in Algorithm 1 with probability 1.

\textit{Proof: See Appendix A.}

\textit{Remark}: It is worth noting that the projection problem (AP) is actually an interesting problem with hidden physical meanings. The projection step is equivalent to finding the closest vector in the feasible set to the desired one, which is also mentioned in \cite{de2010code}.

(OBP-AC-SDP) is actually a special case of the general results in \cite{huang2010rank}, where existence of rank 1 optimal solution is guaranteed. However, in this paper, we extend the results to strong duality with probability 1.

\textit{Theorem 4 (Strong duality of (OBP-AC-SDP))}: Suppose the optimal solution for (OBP-AC-SDP) is $\{\mathbf{W}_{k}^{*}\}_{k=1}^{K}$, then $\textrm{rank}(\mathbf{W}_{k}^{*})=1, \forall k=1,...,K$ with probability 1.

\textit{Proof: See Appendix B.}


\textit{Remark }: In the literature, not all SDR problems have rank 1 characterizations and even if some problems are shown to have rank 1 solution in the numerical sense, it is still highly challenging to find an analytical support \cite{luo2010semidefinite}. In general, the rank 1 justification for SDR problems are case specific and the rank 1 property from one problem cannot be directly generalized to another one. We summarize the difference between our work and the existing works \cite{bengtsson1999optimal}, \cite{huang2010rank} and \cite{song2012robust} in Table \ref{summaryRank1}.

\newcommand{\tabincell}[2]{\begin{tabular}{@{}#1@{}}#2\end{tabular}}
\begin{table*}[t]
\centering
\caption{Summary of Rank 1 results}
\centering
\begin{tabular}{|c| c |c|c|}
\hline
Ref. & Problem scenario &  Rank 1 results & Differentiation \\ \hline
\tabincell{c}{Bengtsson, \\\textit{et al.} \cite{bengtsson1999optimal}} & \tabincell{c}{MISO SINR-constrained \\Beamforming} & \tabincell{c}{Strong duality when there \\are only SINR constraints} & \multirow{3}{*}{\tabincell{l}{1. Our problem contains different \\objective (\ref{a2-sdp}) and shaping constraints \\(\ref{a2-sdp-cons}) (due to  \emph{inhomogeneity}) from \cite{bengtsson1999optimal},\\ \cite{huang2010rank} and \cite{song2012robust} (\emph{homogeneous}).\\2. We obtain strong duality with \\probability 1 for our problem. (refer\\ to Theorem 2, 3, 4 and 5)}
}\\ \cline{1-3}
\tabincell{c}{Huang,\\\textit{et al.} \cite{huang2010rank}} & \tabincell{c}{MISO SINR-constrained \\Beamforming with \\shaping constraints} & \tabincell{c}{Existence\footnotemark of rank 1 \\solution  under limited \\shaping constraints \cite{huang2010rank}}& \\ \cline{1-3}
\tabincell{c}{Song,\\\textit{et al.} \cite{song2012robust}} & \tabincell{c}{Robust SINR-constrained \\Beamforming} & \tabincell{c}{Strong duality under two\\ transmit antennas or special \\uncertainty region}& \\ \hline

\end{tabular}

\label{summaryRank1}
\end{table*}

\footnotetext{Existence of rank 1 optimal solution means that there exists a rank 1 optimal solution among all the optimal solutions, which may have various rank profiles. It is considered as a weaker result than strong duality.}

\textit{Theorem 5 (Convergence of Algorithm 1 to the stationary point of (P0)):} The limiting output $\{\breve{\mathbf{w}}_{k}^{J}\}_{k=1}^{K}$ of algorithm 1 is a stationary point of the non-smooth problem (P0) with probability 1. Furthermore, all the intermediate outputs are rank 1 almost surely.

\textit{Proof: See Appendix C.}

\subsection{Implementation Considerations}
In this subsection, we shall present the implementation considerations and complexity analysis for Algorithm 1 as follows.

\textbf{Implementation Considerations}: Note that the proposed algorithm is implemented in a centralized way. In practical networks, all the BSs can be connected to a central controller or one of BSs can act as the central controller through the backhaul. The CP collects all the required information such as CSI $\mathbf{h}_{k}$ and SINR $\mathbf{\gamma}_{k}$ and then computes the beamforming parameters in a centralized manner. Hence, the beamforming solutions could be distributed to each corresponding BSs through the backhaul.

\textbf{Complexity Analysis}: Note that (OBP-SDP), (AP-SDP) and (OBP-AC-SDP) could be solved by SeDuMi by means of an interior point method and they have the same worst-case complexity order $O((K^{3.5}L^{6.5}N^{6.5})\mathrm{log}(1/\varepsilon))$, where $\varepsilon$ represents the accuracy of the solution at the algorithm's termination. Since (OBP-SDP) and (OBP-AC-SDP) are solved only once for initialization and finalization, respectively, the complexity of the proposed algorithm mainly comes from the repeated executions of (AP-SDP). Hence, the worst-case complexity of the proposed algorithm is $O((K^{3.5}L^{6.5}N^{6.5})\mathrm{log}(1/\varepsilon)\mathrm{log}(1/\theta_{\mathrm{min}}))$, where $\theta_{\mathrm{min}}$ represents the accuracy of the approximation to the non-smooth $l_{0}$ norm. This complexity is substantially lower than the brute force search complexity $O(2^{KN})$.

\section{Numerical Results}
\subsection{Setting of the Algorithm Parameters}
The initial value $\theta^{1}$ is set to two times the maximum absolute value of the entries in the vectors $\{\mathbf{w}_{k}^{0}\}_{k=1}^{K}$. The reason is that this value of $\theta$ is virtually like infinity for all the entries in $\{\mathbf{w}_{k}^{0}\}_{k=1}^{K}$. The update of $\theta$ is controlled by $\tau$ and $\eta$. Note that we do not need to wait for the convergence of the internal loop. For a smaller value of $\tau$, more iterations in the internal loop are needed. Depending on the network configuration, the choice of $\tau$ should be empirically between $KLN\sqrt{\gamma}/3$ and $KLN\sqrt{\gamma}$, where $\gamma$ is SINR.
The $\theta$ decreasing factor $\eta$ is typically chosen from $0.7\thicksim0.95$ and it is set to 0.9 in the simulation. A larger $\eta$ requires more iterations and it is more likely to get the global optimal. The stepsize $\mu^{j}$ is chosen to be decreasing with the smooth parameter $\theta^{j}$, which is because the smooth function fluctuates more with smaller $\theta$; therefore smaller $\mu$ should be applied and it is set to $2(\theta^{j})^{2}$. The small perturbation factor $\varrho$ is chosen to be $0.0001$ to have neglecting effect on the gradient. The algorithm will stop once $\theta$ is below a sufficiently small threshold value $\theta_{\textrm{min}}$. The summary of the algorithm parameters is given in Table \ref{para_sum}.

\begin{table}[ht]
\caption{Summary of Algorithm Parameters}
\centering
\begin{tabular}{|c| l|l |}
\hline

\hline
Para. & Function & Typical values \\ \hline
$\theta^{1}$ & initial value of $\theta$ & $(2\sim4)*\mathop{\textrm{max}}_{n,k,l}|w_{n,k,l}^{0}|$ \\ \hline
$\eta$ & $\theta$ decreasing factor & $0.7 \sim 0.95$ \\ \hline
$\theta_{\mathrm{min}}$ & threshold value of $\theta$ & $0.0001\sim0.001$ \\ \hline
$\tau$ & $\theta$ update control &$KLN\sqrt{\gamma}/3 \sim KLN\sqrt{\gamma}$\\ \hline
$\mu^{j}$ & stepsize & $2(\theta^{j})^{2}\sim3(\theta^{j})^{2}$ \\ \hline
$\varrho$ & perturbation factor & $0.0001\sim0.001$\\ \hline
\end{tabular}
\label{para_sum}
\end{table}

\subsection{Impact of parameter $\epsilon$ on the tradeoff between Cooperation and Power Consumption}
For different network configurations, $\epsilon$ should be properly chosen to yield the desired tradeoff. Generally, a smaller $\epsilon$ promotes a more sparse solution, but with potentially high power consumption, while a larger $\epsilon$ takes power consumption as a priority, which may result in more cooperations. It is worth noting that when $\epsilon$ equals $0$, the design problem only focuses on the minimization of the active backhaul links, regardless of the transmit power consumption. For a very large $\epsilon$, the weight of the power consumption ($l_{2}$ norm part) overwhelms the $l_{0}$ norm counterpart; thus the problem reduces to an (OBP) problem.

\subsection{Random Network Model and Baseline Setups}

\textbf{Random Network Model: }
Consider a Poisson Point Process (PPP) model for the BSs and MSs in a 1km*1km square area with the BS density given by $\lambda_{\textrm{BS}}=4/\mathrm{km}^{2}$ and MS density given by $\lambda_{\textrm{MS}}=8/\mathrm{km}^{2}$. Consider that each BS has $L=2$ antennas and each MS has single antenna.
The channel coefficient between the $l$-th antenna of the $n$-th BS and the $k$-th MS is:
\begin{equation}
h_{n,k,l}=\Gamma_{n,k,l}\sqrt{G\beta d_{n,k}^{-\xi}\zeta_{n,k}},
\end{equation}

\noindent where $d_{n,k}$ is the distance between the $n$-th BS and the $k$-th MS. $G$ is the BS antenna power gain, which is assumed to be 9dB. $\xi$ is the path-loss exponent, and $\beta$ is the path-loss constant. $\zeta_{n,k}\sim \mathcal{N}(0\mathrm{dB},8\mathrm{dB})$ is the corresponding log-normal coefficient, which models the large scale fading (shadowing). $\Gamma_{n,k}^{l}\sim \mathcal{NC}(0,1)$ is the complex Gaussian coefficient, which models the small scale fading. For the pathloss model, the 3GPP Long Term Evolution (LTE) standard has been used:
\begin{equation}
PL_{n,k}^{\mathrm{dB}}=148.1+37.6\mathrm{log}_{10}(d_{n,k}^{\mathrm{km}}).
\end{equation}

For each random network configuration, $10^{3}$ runs are used to simulate the PPP model location. For each PPP model location, $10^{3}$ channel realizations are used to simulate the fading channel.

Five baseline schemes are considered for comparison£º
\begin{enumerate}

\item Baseline 1 is a directional cooperation scheme which greedily chooses the links by evaluating the strength of the channel response\cite{waihoito}. A deflation heuristic is used to minimize the selected links.
\item Baseline 2 is an opportunistic cooperation scheme with suitable switching between the full cooperation model and the coordination mode according to the users' SINR constraints \cite{Gesbert2010}.
\item Baseline 3 is a dynamic clustering scheme that dynamically forms user-centric clusters for each MS to mitigate interference \cite{papadogiannis2008dynamic}. A greedy algorithm is used in this scheme by starting from the lowest number of cooperating clusters to the full cooperation scenario.
\item Baseline 4 is an iteratively link removal algorithm in \cite{zhaocoordinated} that iteratively removes the links that correspond to the smallest link transmit power based on the $l_{2}$ norm relaxation.
\item Baseline 5 is a reweighted $l_{1}$ norm minimization algorithm in \cite{zhaocoordinated} that is based on $l_{1}$ norm relaxation. The algorithm solves a series of $l_{1}$-norm minimization problems to obtain the minimum backhaul cooperation solution.
\end{enumerate}

\subsection{Performance comparison with Baselines}
\begin{figure}
\centering
\includegraphics[width=1.0\columnwidth,draft=false]{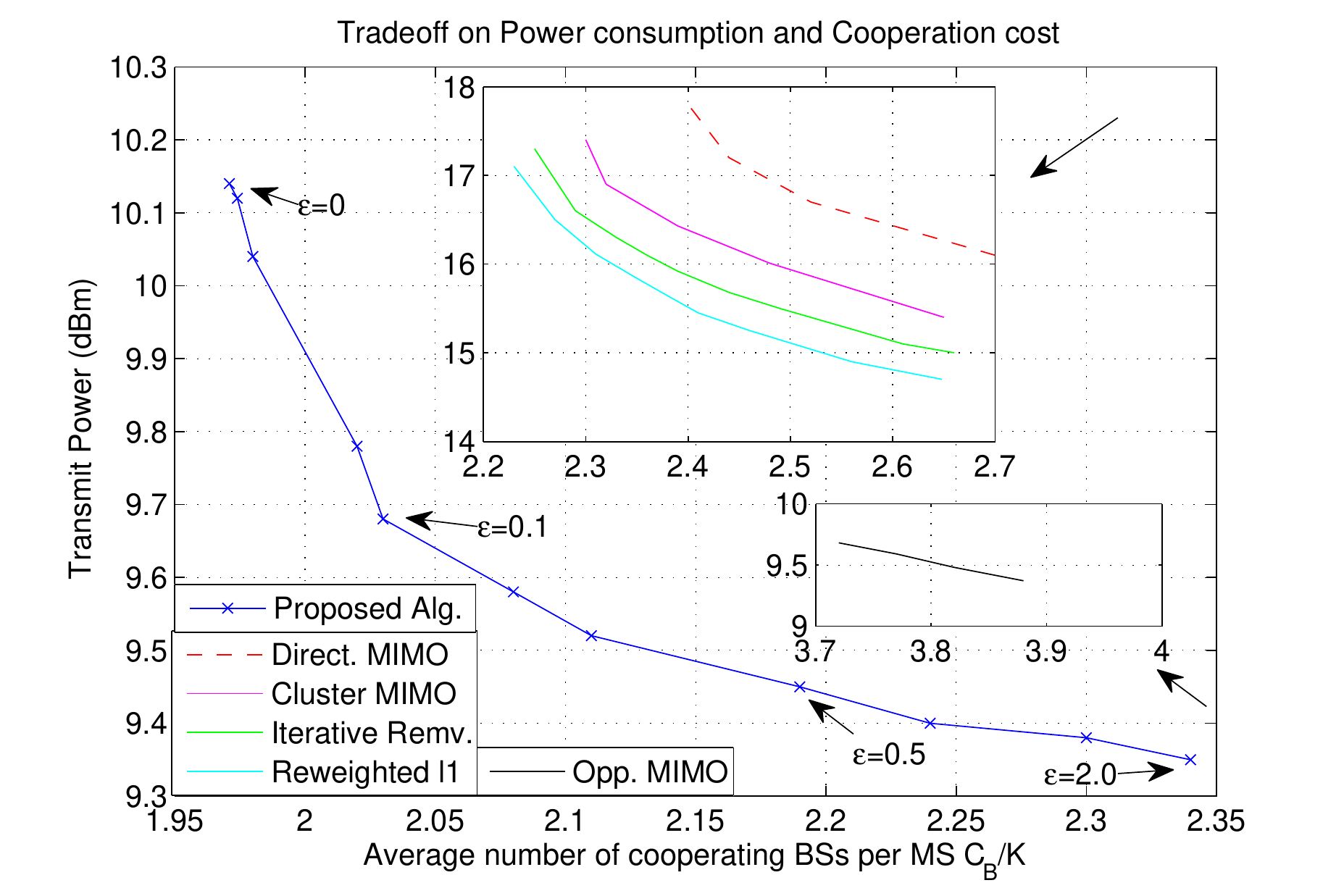}
\caption{Tradeoff on the power consumption and backhaul cost. $L=2$, $\lambda_{\textrm{BS}}=4/\textrm{km}^{2}$, $\lambda_{\textrm{MS}}=8/\textrm{km}^{2}$, area=1km*1km, SINR $\gamma_{k}$ = 15 dB.} \label{fig:fig_tradeoff}
\end{figure}

Figure \ref{fig:fig_tradeoff} investigates the impact of $\epsilon$, which controls the tradeoff between cooperation and power consumption in the proposed algorithm (Note the curves are not plotted together due to the different operating regions of the respective schemes). From Figure \ref{fig:fig_tradeoff}, it is readily observed that the curve of the proposed algorithm is strictly lower than the other baselines. In other words, when fixing the number of cooperating BSs per MS, the proposed algorithm consumes less transmit power than the baselines. When fixing the transmit power, the proposed algorithm requires less number of cooperations than the baselines.

\begin{figure}
\centering
\includegraphics[width=1.0\columnwidth,draft=false]{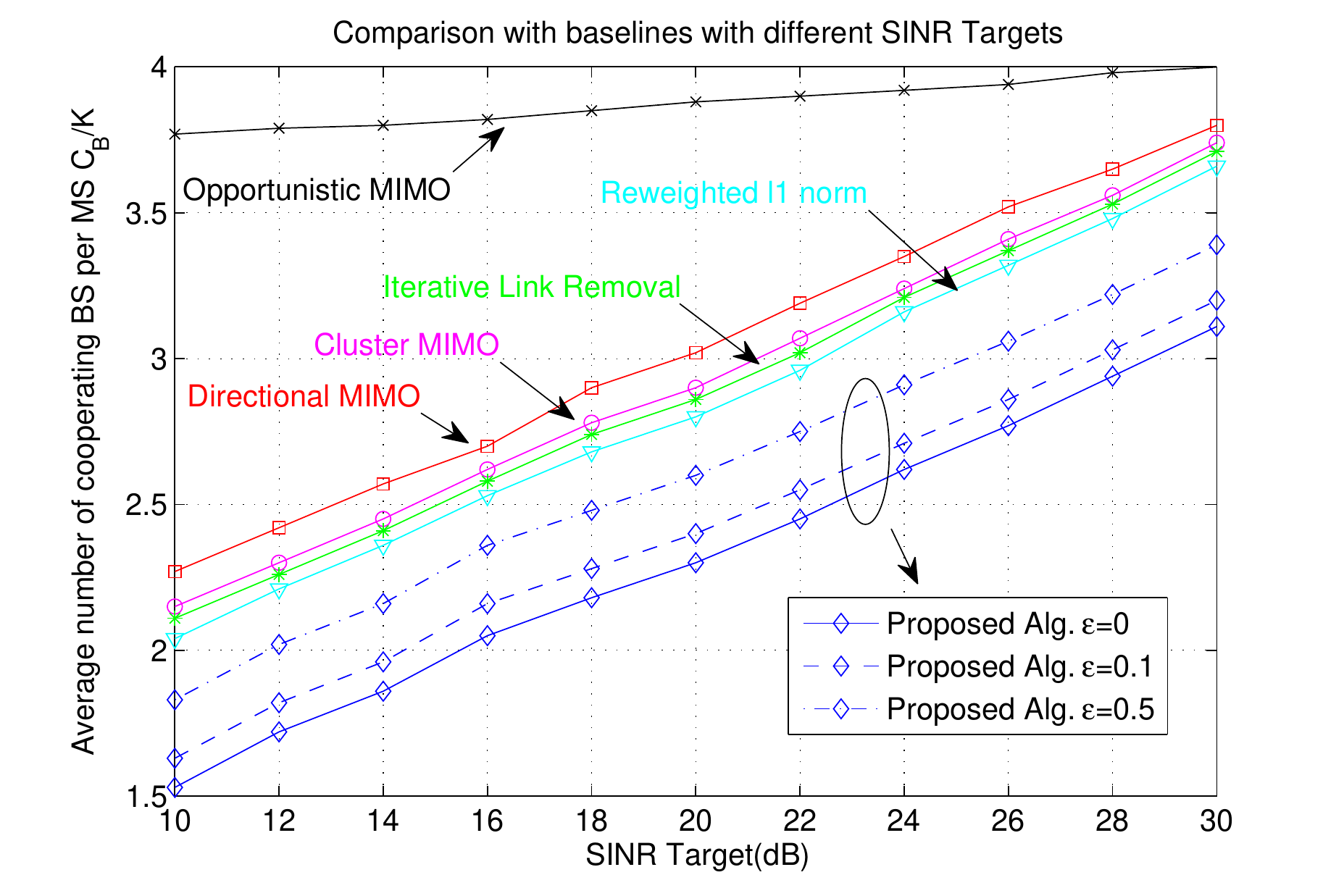}
\caption{Comparison with baselines with different SINRs in terms of Backhaul Cost. $L=2$, $\lambda_{\textrm{BS}}=4/\textrm{km}^{2}$, $\lambda_{\textrm{MS}}=8/\textrm{km}^{2}$ and area=1km*1km.} \label{fig:fig_alg_bas_SINR_density_coop}
\end{figure}

Figure \ref{fig:fig_alg_bas_SINR_density_coop} and Table \ref{pwrComp} illustrate the backhaul cost and power consumption respectively for different schemes versus the SINR requirements $\gamma_{k}$ (assume $\gamma_{k}$ are equal for the users). Three values of $\epsilon$ are considered, $\epsilon=0$, $\epsilon=0.1$ and $\epsilon=0.5$. Specifically, when $\epsilon=0$, the proposed algorithm achieves the minimum backhaul cost at the expense of more power consumption compared with $\epsilon=0.1$ and $\epsilon=0.5$. From Figure \ref{fig:fig_alg_bas_SINR_density_coop}, we observe that the backhaul cost increases as the SINR increases. Moreover, we see that the proposed scheme always have a better performance than the baselines both in terms of the backhaul cost and the power consumption.

\begin{figure}
\centering
\includegraphics[width=1.0\columnwidth,draft=false]{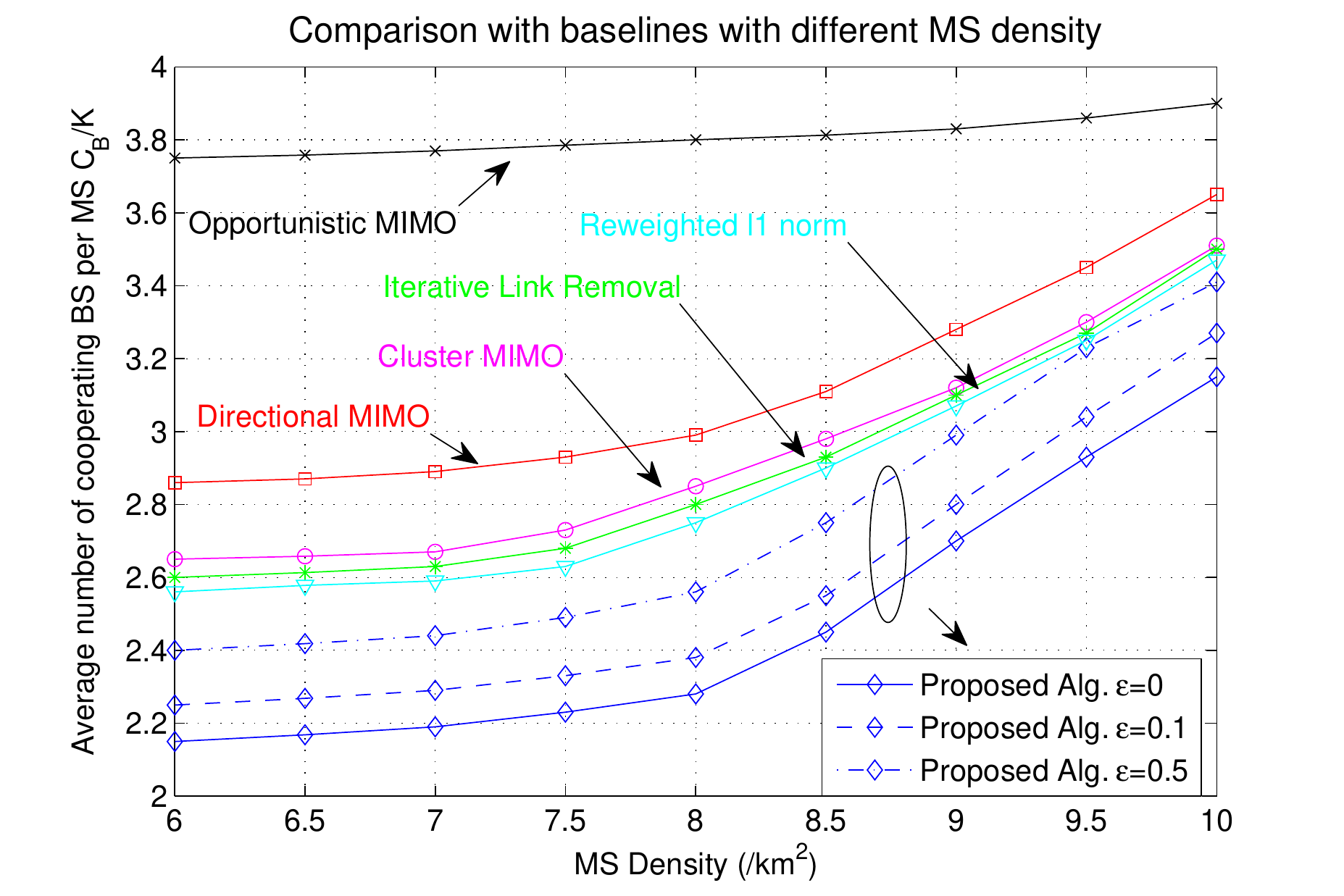}
\caption{Comparison with baselines with different MS density in terms of Backhaul Cost. $L=2$, $\lambda_{\textrm{BS}}=4/\textrm{km}^{2}$, area=1km*1km and SINR $\gamma_{k}$ = 20 dB.} \label{fig:fig_alg_bas_density_coop}
\end{figure}

Figure \ref{fig:fig_alg_bas_density_coop} and Table \ref{pwrComp2} investigate the performance comparison between the proposed algorithm and the baselines as MS density varies in terms of backhaul cost and power consumption, respectively. Similar to the observations we have in Figure \ref{fig:fig_alg_bas_SINR_density_coop} and Table \ref{pwrComp}, the proposed algorithm always has substantial gains over the various baselines in respect to both backhaul cost and power consumption. Moreover, we observe that the backhaul cost increases as the MS density increases.

By jointly processing the signals at the CP, the proposed scheme is able to choose the cooperation links on a larger scale and achieve a better balanced solution. It is worth noting that the proposed algorithm ourperforms the Cluster-MIMO scheme, which means the inter-cluster interference is carefully taken into consideration when jointly designing the asymmetric cooperation and beamforming. It is also noteworthy that the proposed algorithm achieves substantial gains over the reweighted $l_{1}$ norm minimization and iterative link removal algorithms in \cite{zhaocoordinated}. The intuitions behind are as follows. Although $l_{1}$ norm is the best known convex approximation to $l_{0}$ norm, the smoothed $l_{0}$ norm used in this paper is the closest approximation since it is exactly the same as $l_{0}$ norm when the smooth parameter tends to 0. This better approximation and formulation may lead to a better solution over the $l_{1}$ norm based algorithms. On the other hand, the iterative link removal approach proposed is only a \emph{heuristic} algorithm that gradually removes the links purely based on the transmit power.

\newsavebox{\tablebox}
\begin{lrbox}{\tablebox}
\begin{tabular}{|c| c|c|c|c|c|c|c|c|}

\hline
 & \multicolumn{8}{|c|}{Transmit Power (dBm)}\\ \hline
\multirow{2}{*}{\tabincell{c}{SINR\\(dB)}} & \multicolumn{3}{|c|}{Proposed alg.} & \multicolumn{5}{|c|}{Baselines} \\ \cline{2-9}
& $\epsilon=0$ &$0.1$ &$0.5$& B1 & B2 & B3 & B4 & B5\\ \hline
15 & 10.1 & 9.8 & 9.5 & 18.2 & 9.4 & 17.9 & 17.5 & 17.0\\ \hline
20 & 12.0 & 11.7 & 11.5 & 19.9 & 11.3 & 19.6 & 19.3 & 18.9 \\ \hline
25 & 13.8 & 13.4 & 13.1 & 21.4 & 13.0 & 21.1 & 20.8 & 20.5 \\ \hline
30 & 15.4 & 15.0 & 14.6 & 22.8 & 14.4 & 22.4 & 22.1 & 21.8\\ \hline

\end{tabular}
\end{lrbox}

\begin{table}[ht]
\caption{Comparison with baselines in terms of Power consumption with different SINR targets}
\centering

\label{pwrComp}
\scalebox{0.9}{\usebox{\tablebox}}
\end{table}

\newsavebox{\tableboxc}
\begin{lrbox}{\tableboxc}
\begin{tabular}{|c| c|c|c|c|c|c|c|c|}

\hline
 & \multicolumn{8}{|c|}{Transmit Power (dBm)}\\ \hline
\multirow{2}{*}{\tabincell{c}{MS density\\(/$\textrm{km}^{2}$)}} & \multicolumn{3}{|c|}{Proposed alg.} & \multicolumn{5}{|c|}{Baselines} \\ \cline{2-9}
& $\epsilon=0$ &$0.1$ &$0.5$& B1 & B2 & B3 & B4 & B5\\ \hline
6 & 10.1 & 9.7 & 9.6 & 18.5 & 9.5 & 18.0 & 17.7 & 17.3\\ \hline
7 & 11.1 & 10.7 & 10.4 & 19.3 & 10.3 & 18.9 & 18.5 & 18.1 \\ \hline
8 & 12.0 & 11.7 & 11.5 & 19.9 & 11.3 & 19.6 & 19.3 & 18.9 \\ \hline
9 & 12.7 & 12.4 & 12.1 & 20.5 & 12.0 & 20.2 & 20.0 & 19.5\\ \hline
10 & 13.3 & 13.0 & 12.7 & 20.9 & 12.5 & 20.5 & 20.3 & 19.9\\ \hline

\end{tabular}
\end{lrbox}

\begin{table}[ht]
\caption{Comparison with baselines in terms of Power consumption with different MS densities}
\centering

\label{pwrComp2}
\scalebox{0.8}{\usebox{\tableboxc}}
\end{table}




\subsection{Performance comparison with Full search}
Due to the intolerable full search complexity for large problem size, we consider a small random network consists of 3BSs and 3MSs. The SINR threshold $\gamma_{k}$ is set to 20dB for all the users. Table \ref{FullSearch} compares the performance of the proposed algorithm with combinatorial full search in terms of average number of cooperating BSs per MS (denoted as Avg. Coop) and average transmit power (denoted as Avg. Pwr). Three different values of $\epsilon$ are considered, i.e., 0, 0.1, 0.5.

Overall the proposed algorithm performs reasonably close to the optimal case given the tremendous saving in respect to complexity. In this simple setup, full search needs to solve 465 SDP iterations while the proposed scheme only requires 35.2 on average.

\newsavebox{\tableboxb}
\begin{lrbox}{\tableboxb}
\begin{tabular}{|c| c|c |c|c|}
\hline
\multicolumn{5}{|c|}{Performance comparison}\\ \hline
\multirow{2}{*}{$\epsilon$}& \multicolumn{2}{|c|}{Proposed} & \multicolumn{2}{|c|}{Full search}\\ \cline{2-5}
& \tabincell{c}{Avg. Pwr\\(dBm)}& \tabincell{c}{Avg. Coop\\($C_{B}/K$)} & \tabincell{c}{Avg. Pwr\\(dBm)} & \tabincell{c}{Avg. Coop\\($C_{B}/K$)}\\ \hline
0 & 13.78 & 0.92 & 15.43 & 0.80\\ \hline
0.1 & 10.57 & 1.25 & 10.70 & 1.12\\ \hline
0.5 & 7.67 & 1.56 & 7.89 & 1.35\\ \hline
\multicolumn{5}{|c|}{SDP executions}\\ \hline
& \multicolumn{2}{|c|}{35.2} & \multicolumn{2}{|c|}{465}\\ \hline

\hline
\end{tabular}
\end{lrbox}

\begin{table}
\caption{Proposed algorithm v.s. Full search}
\centering

\label{FullSearch}
\scalebox{0.9}{\usebox{\tableboxb}}
\end{table}

\section{Conclusion}
In this work, we propose a novel scheme to reduce the backhaul loading as well as the power consumption in MIMO cellular networks. We model the backhaul cost as the number of direct cooperation links and formulate a combinatorial optimization problem that minimizes the $l_{0}/l_{2}$ norm of the beamforming vectors subject to SINR constraints at all the users. We approximate the non-smooth combinatorial part with a smooth function and relax the non-convex SINR constraints using SDR. A practical and efficient algorithm is introduced to solve the approximated problem. The proposed algorithm guarantees a rank 1 solution with probability 1 and convergence to the stationary point. The effectiveness of the proposed algorithm is demonstrated via extensive simulations.


%

\appendices
\section{Proof of Theorem 3}
Before proving Theorem 3, we introduce the following lemmas.

\textit{Lemma 1}: (Proposition 4, \cite{wiesel2006linear}) If $\mathbf{A}\succeq 0$,$\mathbf{B}\succeq 0$ and $\mathbf{c}$ is in the range of $\mathbf{A}$, then
\begin{equation}
\mathbf{c}^{H}\mathbf{A}^{\dag}\mathbf{c}\geq \mathbf{c}^{H}(\mathbf{A}+\mathbf{B})^{\dag}\mathbf{c},
\end{equation}

\noindent with equality if and only if $\mathbf{B}(\mathbf{A}+\mathbf{B})^{\dag}\mathbf{c}=0$.

\textit{Lemma 2}: If $\mathbf{A}\succeq 0$ and $\mathbf{A}(\mathbf{I}+\mathbf{A})^{-1}\mathbf{c}=\mathbf{0}$, then
\begin{equation}
\mathbf{A}\mathbf{c}=\mathbf{0} \quad \textrm{and} \quad (\mathbf{I}+\mathbf{A})^{-1}\mathbf{c}=\mathbf{c}
\end{equation}
\textit{Proof}: Since $(\mathbf{I}+\mathbf{A})$ is a full rank matrix, there always exists a $\mathbf{y}$ such that $(\mathbf{I}+\mathbf{A})\mathbf{y}=\mathbf{c}, \forall \mathbf{c}$. Therefore, $\mathbf{A}(\mathbf{I}+\mathbf{A})^{-1}(\mathbf{I}+\mathbf{A})\mathbf{y}=\mathbf{0}$, i.e. $\mathbf{A}\mathbf{y}=\mathbf{0}$. This means $\mathbf{y}\in \mathcal{N}(\mathbf{A})$ and $\mathbf{c}=\mathbf{y}=(\mathbf{I}+\mathbf{A})^{-1}\mathbf{c}$. $\Box$

\textit{Proof of Theorem 3: }We first relax the constraint $\mathrm{Tr}(\mathbf{I}_{00}\tilde{\mathbf{W}}_{k})=1$ in (AP-SDP) to $\mathrm{Tr}(\mathbf{I}_{00}\tilde{\mathbf{W}}_{k})\leq1$, resulting in the relaxed problem (AP-SDP-Relaxed):
\begin{align}
\mathop{\textrm{min}}_{\{\tilde{\mathbf{W}}_{k}\}^{K}_{k=1}} \quad & \sum_{k=1}^{K}\mathrm{Tr}(\mathbf{A}_{k}\tilde{\mathbf{W}}_{k})  \\
\textrm{s.t.} \qquad & \mathrm{Tr}(\tilde{\mathbf{H}}_{k}\tilde{\mathbf{W}}_{k})-\gamma_{k}\mathop{\sum}_{m\neq k}\mathrm{Tr}(\tilde{\mathbf{H}}_{k}\tilde{\mathbf{W}}_{m})-\gamma_{k}\sigma_{k}^{2} \geq 0  \nonumber\\ &
\mathbf{Tr}(\mathbf{I}_{00}\tilde{\mathbf{W}}_{k})\leq1,\quad\forall k \nonumber\\
& \tilde{\mathbf{W}}_{k}\succeq \mathbf{0}, \quad\forall k\nonumber
\end{align}
We shall first prove that the SDR of (AP-SDP-Relaxed) is always tight given the sufficient condition that $\mathbf{H}_{k}\mathbf{\bar{w}}_{k}\neq \mathbf{0}, \forall k$.
Let the optimal value of (AP-SDP) be $p^{*}$ and the optimal value of (AP-SDP-Relaxed) be $d^{*}$. Then $p^{*}\geq d^{*}$. The lagrangian dual problem of (AP-SDP-Relaxed) is shown below. $\lambda_{k}$, $\varphi_{k}$ and $\mathbf{Z}_{k}$ are the lagrangian multipliers for the SINR inequality, the trace inequality and the PSD constraint, respectively.\\
\\ \indent (AP-SDP-Relaxed-Dual)
\begin{align}
\mathop{\textrm{max}}_{\{\lambda_{k},\varphi_{k}\}^{K}_{k=1}} \quad & \sum_{k=1}^{K}(\lambda_{k}\gamma_{k}\sigma_{k}^{2}-\varphi_{k})  \\
\textrm{s.t.} \qquad & \mathbf{Z}_{k}=\mathbf{A}_{k}-\lambda_{k}\tilde{\mathbf{H}}_{k}+\sum_{m\neq k}\lambda_{m}\gamma_{m}\tilde{\mathbf{H}}_{m}+\varphi_{k}\mathbf{I}_{00}\succeq0\nonumber\\
& \lambda_{k} \geq0, \quad\forall k\nonumber\\
& \varphi_{k} \geq0, \quad\forall k\nonumber
\end{align}
$\mathbf{Z}_{k}$ can be written in the following matrix form:
\begin{align}
&\mathbf{Z}_{k}=\\\nonumber
&\left[\begin{array}{cccc}{\bar{\mathbf{w}}_{k}^{H}\bar{\mathbf{w}}_{k}+\varphi_{k}}&{-\bar{\mathbf{w}}_{k}^{H}}
  \\{-\bar{\mathbf{w}_{k}}}&{\mathbf{I}+\sum_{m\neq k}\lambda_{m}\gamma_{m}\mathbf{H}_{m}-\lambda_{k}\mathbf{H}_{k}}\end{array}\right]\succeq0.
\end{align}

If $\varphi_{k}>0$, then by Schur complement
\begin{align}
\mathbf{A}_{k}+\varphi_{k}\mathbf{I}_{00}=
\left[\begin{array}{cccc}{\bar{\mathbf{w}}_{k}^{H}\bar{\mathbf{w}}_{k}+\varphi_{k}}&{-\bar{\mathbf{w}}_{k}^{H}}
  \\{-\bar{\mathbf{w}_{k}}}&{\mathbf{I}}\end{array}\right]\succ0.
\end{align}

\noindent since $\bar{\mathbf{w}}_{k}^{H}\bar{\mathbf{w}}_{k}+\varphi_{k}-\bar{\mathbf{w}}_{k}^{H}\bar{\mathbf{w}}_{k}>0$. Thus $\mathbf{A}_{k}+\varphi_{k}\mathbf{I}_{00}+\sum_{m\neq k}\lambda_{m}\gamma_{m}\tilde{\mathbf{H}}_{m}$ is also a full rank matrix. By KKT conditions, we know,
\begin{equation}
\label{Zl}
\mathbf{A}_{k}+\sum_{m\neq k}\lambda_{m}\gamma_{m}\tilde{\mathbf{H}}_{m}+\varphi_{k}\mathbf{I}_{00}-\lambda_{k}\tilde{\mathbf{H}}_{k}-\mathbf{Z}_{k}=\mathbf{0},
\end{equation}
\begin{equation}
\label{KKT}
\tilde{\mathbf{W}}_{k}\mathbf{Z}_{k}=\mathbf{0}.
\end{equation}

Premultiplying the two sides of (\ref{Zl}) by $\tilde{\mathbf{W}}_{k}$, and making use of (\ref{KKT}), we get
\begin{equation}
\tilde{\mathbf{W}}_{k}(\mathbf{A}_{k}+\sum_{m\neq k}\lambda_{m}\gamma_{m}\tilde{\mathbf{H}}_{m}+\varphi_{k}\mathbf{I}_{00})=\lambda_{k}\tilde{\mathbf{W}}_{k}\tilde{\mathbf{H}}_{k}.
\end{equation}

Now the following relation holds:
\begin{align}
\label{rank}
\textrm{rank}(\tilde{\mathbf{W}}_{k})=& \textrm{rank}(\tilde{\mathbf{W}}_{k}(\mathbf{A}_{k}+\sum_{m\neq k}\lambda_{m}\gamma_{m}\tilde{\mathbf{H}}_{m}+\varphi_{k}\mathbf{I}_{00}))\\
= & \textrm{rank}(\lambda_{k}\tilde{\mathbf{W}}_{k}\tilde{\mathbf{H}}_{k})
\\ \leq & \textrm{min}\{ \textrm{rank}(\tilde{\mathbf{H}}_{k}),\textrm{rank}(\tilde{\mathbf{W}}_{k})\},\label{min_rank}
\end{align}

\noindent where (\ref{rank}) is due to $\mathbf{A}_{k}+\sum_{m\neq k}\lambda_{m}\gamma_{m}\tilde{\mathbf{H}}_{m}+\varphi_{k}\mathbf{I}_{00}$ being a full rank matrix. and (\ref{min_rank}) follows from a basic rank inequality property \cite{horn1990matrix}. Because $\tilde{\mathbf{H}}_{k}$ is rank 1, we obtain $\textrm{rank}(\tilde{\mathbf{W}}_{k})\leq1$. Since $\tilde{\mathbf{W}}_{k}\neq 0$, the rank of $\tilde{\mathbf{W}}_{k}$ must be 1.

For $\varphi_{k}=0$, because $\mathbf{Z}_{k}\succeq0$, by Generalized Schur Complement (for a singular bottom right corner matrix), we have
\begin{equation}
\bar{\mathbf{w}}_{k}^{H}\bar{\mathbf{w}}_{k}+\varphi_{k}-\bar{\mathbf{w}}_{k}^{H}(\mathbf{I}+\sum_{m\neq k}\lambda_{m}\gamma_{m}\mathbf{H}_{m}-\lambda_{k}\mathbf{H}_{k})^{\dag}\bar{\mathbf{w}_{k}}\geq0,
\end{equation}
\begin{equation}
\bar{\mathbf{w}}_{k} \in \mathfrak{R}(\mathbf{I}+\sum_{m\neq k}\lambda_{m}\gamma_{m}\mathbf{H}_{m}-\lambda_{k}\mathbf{H}_{k})
\end{equation}
\begin{equation}
\mathbf{I}+\sum_{m\neq k}\lambda_{m}\gamma_{m}\mathbf{H}_{m}-\lambda_{k}\mathbf{H}_{k}\succeq0.
\end{equation}

Therefore, by \textit{Lemma 1},
\begin{align}
\label{inequality_mu}
\bar{\mathbf{w}}_{k}^{H}\bar{\mathbf{w}}_{k} \geq &\bar{\mathbf{w}}_{k}^{H}(\mathbf{I}+\sum_{m\neq k}\lambda_{m}\gamma_{m}\mathbf{H}_{m}-\lambda_{k}\mathbf{H}_{k})^{\dag}\bar{\mathbf{w}_{k}} \\ \nonumber & \geq \bar{\mathbf{w}}_{k}^{H}(\mathbf{I}+\sum_{m\neq k}\lambda_{m}\gamma_{m}\mathbf{H}_{m})^{-1}\bar{\mathbf{w}_{k}} \quad \textcircled{1} \\ \nonumber
\bar{\mathbf{w}}_{k}^{H}\bar{\mathbf{w}}_{k} \geq& \bar{\mathbf{w}}_{k}^{H}(\mathbf{I}+\sum_{m\neq k}\lambda_{m}\gamma_{m}\mathbf{H}_{m})^{-1}\bar{\mathbf{w}_{k}}. \quad \textcircled{2}
\end{align}

The conditions for the equality to hold in $\textcircled{1}$ and $\textcircled{2}$ are
\begin{equation}
\label{Hk}
\lambda_{k}\mathbf{H}_{k}(\mathbf{I}+\sum_{m\neq k}\lambda_{m}\gamma_{m}\mathbf{H}_{m})^{-1}\bar{\mathbf{w}}_{k}=0
\end{equation}
\begin{equation}
\label{Hm}
(\sum_{m\neq k}\lambda_{m}\gamma_{m}\mathbf{H}_{m})(\mathbf{I}+\sum_{m\neq k}\lambda_{m}\gamma_{m}\mathbf{H}_{m})^{-1}\bar{\mathbf{w}}_{k}=0
\end{equation}

Note that if (\ref{Hm}) holds, then by \textit{Lemma 2}, $(\mathbf{I}+\sum_{m\neq k}\lambda_{m}\gamma_{m}\mathbf{H}_{m})^{-1}\bar{\mathbf{w}}_{k}=\bar{\mathbf{w}}_{k}$ and $(\sum_{m\neq k}\lambda_{m}\gamma_{m}\mathbf{H}_{m})\bar{\mathbf{w}}_{k}=0$. Thus, (\ref{Hk}) can be written as $\lambda_{k}\mathbf{H}_{k}\bar{\mathbf{w}}_{k}=0$. If the equality in $\textcircled{2}$ holds (i.e., (\ref{Hm})), then (\ref{Hk}) has to hold. Given the sufficient condition that $\mathbf{H}_{k}\mathbf{\bar{w}}_{k}\neq\mathbf{0}$, $\lambda_{k}$ has to be 0. Then $\mathbf{Z}_{k}=\mathbf{A}_{k}+\sum_{m\neq k}\lambda_{m}\gamma_{m}\tilde{\mathbf{H}}_{m}$ and $\mathrm{rank}(\mathbf{Z}_{k})=N$; thus rank($\tilde{\mathbf{W}}_{k}$)=1.

If (\ref{Hm}) does not hold, then $\mathbf{A}_{k}+\sum_{m\neq k}\lambda_{m}\gamma_{m}\tilde{\mathbf{H}}_{m}$ is a full rank matrix. Using the same relation in (\ref{rank}), we have rank($\tilde{\mathbf{W}}_{k}$)=1.

Summing the statements above, the solution of (AP-SDP-Relaxed) must be rank 1. As a matter of fact, the sufficient condition $\mathbf{H}_{k}\mathbf{\bar{w}}_{k}\neq\mathbf{0}$ holds with probability 1 in Algorithm 1. Since $\bar{\mathbf{w}}_{k}=\mathbf{w}_{k}^{j-1}+\mu^{j}\delta_{k}-\mu^{j}\varsigma\mathbf{w}_{k}^{j-1}$, if $\mathbf{H}_{k}\mathbf{\bar{w}}_{k}=\mathbf{0}$ then $\mu^{j}\varsigma\mathbf{H}_{k}\mathbf{w}_{k}^{j-1}=\mathbf{H}_{k}(\mathbf{w}_{k}^{j-1}+\mu^{j}\delta_{k})$. However, the right hand side is predetermined and $\mathbf{H}_{k}\mathbf{w}_{k}^{j-1}\neq\mathbf{0}$, the above equality holds with probability 0 when $\varsigma$ is randomly chosen.When $\mu^{j}\rightarrow0$, the right hand side reduces to $\mathbf{H}_{k}\mathbf{w}_{k}^{j-1}$, which is nonzero as well. Therefore, the sufficient condition holds with probability 1 in Algorithm 1.

Also note that $\tilde{\mathbf{W}}_{k}\succeq0$ and $\mathrm{Tr}(\mathbf{I}_{00}\tilde{\mathbf{W}}_{k})\leq1$, then $\tilde{\mathbf{W}}_{k}$ should behave in the following structure:
\begin{equation}
\tilde{\mathbf{W}}_{k}=\left[\begin{array}{cccc}{z^{2}}&{z\mathbf{w}_{k}^{H}}
  \\{z\mathbf{w}_{k}}&{\mathbf{w}_{k}\mathbf{w}_{k}^{H}}\end{array}\right]
\end{equation}

\noindent with $0<z\leq1$. Thus, from the KKT conditions $\mathbf{Z}_{k}\tilde{\mathbf{W}}_{k}=\mathbf{0}$ and $\varphi_{k}(\mathrm{Tr}(\mathbf{I}_{00}\tilde{\mathbf{W}}_{k})-1)=0$, we have
\begin{equation}
\label{KKTpsd2}
(\mathbf{I}+\sum_{m\neq k}\lambda_{m}\gamma_{m}\mathbf{H}_{m}-\lambda_{k}\mathbf{H}_{k})\mathbf{w}_{k}=z\mathbf{\bar{w}}_{k},
\end{equation}
\begin{equation}
\label{KKTpsd1}
z^{2}(\mathbf{\bar{w}}_{k}^{H}\mathbf{\bar{w}}_{k}+\varphi_{k})=z\mathbf{\bar{w}}_{k}^{H}\mathbf{w}_{k}
\end{equation}
\begin{equation}
\label{KKTphi}
\varphi_{k}(z-1)=0.
\end{equation}

Thus if $\varphi_{k}>0$, by (\ref{KKTphi}), $z=1$. If $\varphi_{k}=0$, from (\ref{KKTpsd1}) we know $z^{2}\mathbf{\bar{w}}_{k}^{H}\mathbf{\bar{w}}_{k}=z\mathbf{\bar{w}}_{k}^{H}\mathbf{w}_{k}$. Note $\mathbf{\bar{w}}_{k}^{H}\mathbf{w}_{k}$ is a positive scaler, which can be easily observed by premultiplying $\mathbf{w}_{k}$ to (\ref{KKTpsd2}).

$\tilde{\mathbf{W}}_{k}$ is a feasible solution to (AP-SDP-Relaxed) $\forall 0<z\leq1$, since the choice of $z$ will not affect the SINR constraint and positive semidefinite requirement. By generalized Schur complement, $z^{2}\geq(z\mathbf{w}_{k}^{H})(\mathbf{w}_{k}\mathbf{w}_{k}^{H})^{\dag}(z\mathbf{w}_{k}))$, which can be simplified as $1\geq\mathbf{w}_{k}^{H}(\mathbf{w}_{k}\mathbf{w}_{k}^{H})^{\dag}\mathbf{w}_{k}$. The equality holds and obviously $\forall 0<z\leq1$.

In (AP-SDP-Relaxed), the objective can be written as
\begin{equation}
\mathrm{Tr}(\mathbf{A}_{k}\tilde{\mathbf{W}}_{k})=z^{2}\mathbf{\bar{w}}_{k}^{H}\mathbf{\bar{w}}_{k}-z\mathbf{\bar{w}}_{k}^{H}\mathbf{w}_{k}
-z\mathbf{w}_{k}^{H}\mathbf{\bar{w}}_{k}+\mathbf{w}_{k}^{H}\mathbf{w}_{k},
\end{equation}

Since $z^{2}\mathbf{\bar{w}}_{k}^{H}\mathbf{\bar{w}}_{k}=z\mathbf{\bar{w}}_{k}^{H}\mathbf{w}_{k}
=z\mathbf{w}_{k}^{H}\mathbf{\bar{w}}_{k}$, then $\mathrm{Tr}(\mathbf{A}_{k}\tilde{\mathbf{W}}_{k})=
-z\mathbf{w}_{k}^{H}\mathbf{\bar{w}}_{k}+\mathbf{w}_{k}^{H}\mathbf{w}_{k}=-z^{2}\mathbf{\bar{w}}_{k}^{H}\mathbf{\bar{w}}_{k}
+\mathbf{w}_{k}^{H}\mathbf{w}_{k}
\geq -\mathbf{\bar{w}}_{k}^{H}\mathbf{\bar{w}}_{k}
+\mathbf{w}_{k}^{H}\mathbf{w}_{k},$ with the equality holding when $z=1$. Therefore, $z$ must be 1 for the optimal solution. Thus, $\tilde{\mathbf{W}}_{k}$ is also the optimal solution for (AP-SDP) and $p^{*}
=d^{*}$, which implies that the problems (AP-SDP) and (AP-SDP-Relaxed) are equivalent. Therefore, the optimal solution of (AP-SDP) must be rank 1. $\Box$

\section{Proof of Theorem 4}

\textit{Proof: }The lagrangian dual problem of (OBP-AC-SDP) is shown below. $\lambda_{k}$, $\varphi_{k}$ and $\mathbf{Z}_{k}$ are the lagrangian multipliers for the SINR inequality, the trace equality and the PSD constraint, respectively.\\
\indent (OBP-AC-SDP-Dual)
\begin{align}
\label{OBP-AC-DUAL}
\mathop{\textrm{max}}_{\{\lambda_{k},\varphi_{k}\}^{K}_{k=1}} \quad & \sum_{k=1}^{K}\lambda_{k}\gamma_{k}\sigma_{k}^{2}  \\
\textrm{s.t.} \qquad & \mathbf{Z}_{k}=\mathbf{I}-\lambda_{k}\mathbf{H}_{k}+\sum_{m\neq k}\lambda_{m}\gamma_{m}\mathbf{H}_{m}+\varphi_{k}\mathbf{M}_{k}\succeq0 \nonumber\\
& \lambda_{k} \geq0, \quad\forall k\nonumber
\end{align}

Note that the problem (OBP-AC-SDP-Dual) and (OBP-SDP-Dual) (which is the OBP with full cooperation) have the same objective function and only differs in the term $\varphi_{k}\mathbf{M}_{k}$ in the constraints. Suppose (OBP-SDP-Dual) and (OBP-AC-SDP-Dual) have the same predetermined parameters $\{\mathbf{H}_{k}\}_{k=1}^{K}$ and $\{\gamma_{k}\}_{k=1}^{K}$, the optimal solution of (OBP-SDP-Dual) is $\{\lambda_{k}^{*}\}_{k=1}^{K}$ and the optimal solution of (OBP-AC-SDP-Dual) is $\{\tilde{\lambda}_{k}^{*}\}_{k=1}^{K}$.
The feasible set of (\ref{OBP-AC-DUAL}) is equivalent to
\begin{equation}
\mathbf{I}-\lambda_{k}\mathbf{H}_{k}+\sum_{m\neq k}\lambda_{m}\gamma_{m}\mathbf{H}_{m}\succeq-\varphi_{k}\mathbf{M}_{k}
\end{equation}

Then if $\varphi_{k}\leq0$, the feasible set will be smaller or equal to the feasible set in (\ref{A1-SDP-Dual}). Thus the optimal objective value of the solution $\{\tilde{\lambda}_{k}^{*}\}_{k=1}^{K}$ will be smaller or equal to the objective value of $\{\lambda_{k}^{*}\}_{k=1}^{K}$.

Notice that $\mathbf{M}_{k}$ is a predetermined diagonal matrix with diagonal elements being 1 or 0. Let the null space vector of $\mathbf{I}-\lambda_{k}^{*}\mathbf{H}_{k}+\sum_{m\neq k}\lambda_{m}^{*}\gamma_{m}\mathbf{H}_{m}$ be $\mathbf{y}_{k}$. Thus, $\mathbf{y}_{k}$ depends on the i.i.d. channel realizations $\mathbf{H}_{k}$. From this fact, $\mathbf{y}_{k}$ is continuously distributed in the $LN$-dimensional space and consequently, the probability that $\mathbf{y}_{k}$ falls in the null space of $\mathbf{M}_{k}$ is zero, i.e., $\mathbf{M}_{k}\mathbf{y}_{k}=\mathbf{0}$ is with probability 0. The case of $\mathbf{M}_{k}=\mathbf{O}$ is degenerated to (\ref{A1-SDP-Dual}).

If $\varphi_{k}>0$, under the condition that $\mathbf{M}_{k}\mathbf{y}_{k}=\mathbf{0}$ with probability 0, then the matrix $\mathbf{I}-\lambda_{k}^{*}\mathbf{H}_{k}+\sum_{m\neq k}\lambda_{m}^{*}\gamma_{m}\mathbf{H}_{m}+\varphi_{k}\mathbf{M}_{k}\succ0$ with probability 1. Thus, $\exists\vartheta>0$ such that
\begin{equation}
\mathbf{I}-\lambda_{k}^{*}\mathbf{H}_{k}+\sum_{m\neq k}\lambda_{m}^{*}\gamma_{m}\mathbf{H}_{m}+\varphi_{k}\mathbf{M}_{k}-\vartheta\mathbf{H}_{k}\succeq0
\end{equation}

Then $\tilde{\lambda}_{k}^{*}=\lambda_{k}^{*}+\vartheta>\lambda_{k}^{*}$ and this $\tilde{\lambda}_{k}^{*}$ is still a feasible solution of (\ref{OBP-AC-DUAL}), but with a larger objective value than $\{\lambda_{k}^{*}\}_{k=1}^{K}$.

Therefore, $\varphi_{k}>0$ when achieving the optimal solution. Then $\mathbf{I}+\sum_{m\neq k}\lambda_{m}\gamma_{m}\mathbf{H}_{m}+\varphi_{k}\mathbf{M}_{k}$ is a full rank matrix. Thus $\mathrm{rank}(\mathbf{I}+\sum_{m\neq k}\lambda_{m}\gamma_{m}\mathbf{H}_{m}+\varphi_{k}\mathbf{M}_{k}-\lambda_{k}\mathbf{H}_{k})\geq N-1$. Then by the KKT condition that $\mathbf{Z}_{k}\mathbf{W}_{k}=\mathbf{O}$, $\mathrm{rank}(\mathbf{W}_{k})\leq1$. Since $\mathbf{W}_{k}\neq 0$, rank($\mathbf{W}_{k}$)=1 with probability 1. $\Box$

\section{Proof of Theorem 5}
\textit{Proof}:
Let $\mathbf{x}=[\mathbf{w}_{1}^{T}\, \mathbf{w}_{2}^{T}\, ...\, \mathbf{w}_{K}^{T}]^{T}$ and $\{\mathbf{x}^{j}\}$ be the intermediate outputs of algorithm 1 in the $j$-th iteration of the smoothing parameter $\theta$.

Let $\mathcal{C}$ denote the feasible convex set of (P2$(\theta)$), shown at the top of the next page.
\begin{figure*}[t]
\begin{equation}
\mathcal{C}=\Big{\{}\mathbf{x}=[\mathbf{w}_{1}^{T}\, \mathbf{w}_{2}^{T}\, ...\, \mathbf{w}_{K}^{T}]^{T}\Big{|}\mathrm{Tr}(\mathbf{H}_{k}\mathbf{W}_{k})-\gamma_{k}\mathop{\sum}_{m\neq k}\mathrm{Tr}(\mathbf{H}_{k}\mathbf{W}_{m})-\gamma_{k}\sigma_{k}^{2} \geq 0,\ \textrm{and} \left[\begin{array}{cccc}{\mathbf{W}_{k}}&{\mathbf{w}_{k}}\\{\mathbf{w}_{k}^H}&{1}\end{array}\right]\succeq\mathbf{0},  \quad \forall k=1,...,K\Big{\}}
\end{equation}
\hrulefill
\end{figure*}

Since $g(\cdot,\theta)$ (defined in Section III. A.) is a smooth function, a sequence of feasible points $\{\mathbf{x}^{j}\}$ in $\mathcal{C}$ will be generated using the projected gradient method:
\begin{equation}
\mathbf{x}^{j+1}=P_{\mathcal{C}}(\mathbf{x}^{j}-\mu^{j}\nabla g(\mathbf{x}^{j})+e(\mathbf{x}^{j},\mu^{j}))
\end{equation}

\noindent for solving
\begin{equation}
\mathop{\textrm{min}}_{\mathbf{x}\in \mathcal{C}} \quad  g(\mathbf{x},\theta),
\end{equation}

\noindent where $\mu^{j}$ is the step size, $e(\mathbf{x}^{j},\mu^{j})$ is the small perturbation tending 0 when $\mu^{j}\rightarrow0$ and $P_{\mathcal{C}}$ denotes the projection onto the set $\mathcal{C}$.

If the sequence $\{\mathbf{x}^{j}\}$ satisfies
\begin{equation}
\mathop{\textrm{lim}}_{j\rightarrow\infty}\|\nabla g(\mathbf{x}^{j},\theta)^{T}(\mathbf{x}^{j+1}-\mathbf{x}^{j})\|=0,
\end{equation}

\noindent then $\mathbf{x}^{j+1}$ is a stationary point \cite{rinaldimathematical}.

Denote $J=\{j|\theta^{j+1}=\eta\theta^{j}\}$. If $J$ is finite, then there exists an integer $\bar{j}$ such that for all $j>\bar{j}$,
\begin{equation}
\label{gradient:stationary}
\|\nabla g(\mathbf{x}^{j},\theta^{j})^{T}(\mathbf{x}^{j+1}-\mathbf{x}^{j})\|\geq\tau\theta^{j},
\end{equation}

\noindent and $\theta^{k}=\bar{\theta}$ for all $j\geq\bar{j}$ in the $\theta$ updating step of the proposed algorithm. (Note $\|\sum_{k=1}^{K}\mathbf{\delta}_{k}(\mathbf{w}_{k}^{j}-\mathbf{w}_{k}^{j-1})\|$ is equivalent to $\|\nabla g(\mathbf{x}^{j},\theta^{j})^{T}(\mathbf{x}^{j+1}-\mathbf{x}^{j})\|$.) We know from above that when solving the smooth optimization problem over a convex set
\begin{equation}
\mathop{\textrm{min}}_{\mathbf{x}\in \mathcal{C}} \quad g(\mathbf{x},\bar{\theta}),
\end{equation}

\noindent the sequence of generated feasible points $\{\mathbf{x}^{j}\}$ will converge to a stationary point\footnote{Note that because $e(\mathbf{w}_{k}^{j-1},\mu^{j})\rightarrow 0$ when $\mu^{j}\rightarrow0$, this small perturbation will not affect the convergence of the projection gradient algorithm, as shown in \cite{solodov1997convergence} and \cite{luo1993error}.} and thus satisfies
\begin{equation}
\mathop{\textrm{lim}}_{j\rightarrow\infty}\|\nabla g(\mathbf{x}^{j},\bar{\theta})^{T}(\mathbf{x}^{j+1}-\mathbf{x}^{j})\|=0,
\end{equation}

\noindent which contradicts (\ref{gradient:stationary}). This shows that $J$ must be infinite and $\mathop{\textrm{lim}}_{j\rightarrow\infty}\theta^{j}=0$.

Since $J$ is infinite, we can assume that $J=\{j_{0},j_{1},...\}$ with $j_{0}<j_{1}<...$. Then we have
\begin{equation}
\mathop{\textrm{lim}}_{i\rightarrow\infty}\|\nabla g(\mathbf{x}^{j_{i}},\theta^{j_{i}})^{T}(\mathbf{x}^{j_{i}+1}-\mathbf{x}^{j_{i}})\|\leq\tau\mathop{\textrm{lim}}_{i\rightarrow\infty}\theta^{j_{i}}=0.
\end{equation}

Let $\mathbf{x}^{J}$ be an accumulation point of $\{\mathbf{x}^{j_{i}+1}\}$, then $\mathbf{x}^{J}$ is a stationary point of (P2$(\theta)$) when $\theta$ tends to 0.

From Theorem 2 and 3, we know that (OBP-SDP) has strong duality and (AP-SDP) have strong duality with probability 1. Therefore, the overall SDPs in Algorithm 1 have strong duality with probability 1, which implies the SDR of (P1$(\theta)$) is tight with probability 1. Thus, the stationary point of (P2$(\theta)$) is the stationary point of (P1$(\theta)$), which is also the stationary point of the original non-smooth problem (P0) by Definition 1. Since $l_{0}$ norm is size-insensitive, an (OBP-AC-SDP) deactivating the zero entries has to be solved to obtain the stationary point with minimum power, $\breve{\mathbf{x}}^{J}$. From Theorem 4, we know that (OBP-AC-SDP) has strong duality with probability 1. Thus, this $\breve{\mathbf{x}}^{J}$ is the stationary point with minimum required power of the original non-smooth problem (P0). $\Box$




\ifCLASSOPTIONcaptionsoff
  \newpage
\fi



%

\bibliography{citation}

%








\end{document}